\documentclass[aps,prc,twocolumn,superscriptaddress,showpacs,floatfix,nofootinbib]{revtex4}
\usepackage{amsmath,graphicx}

\newcommand{\mean}[1]{\left\langle #1 \right\rangle}
\newcommand{\smean}[1]{\langle #1 \rangle}
\newcommand{\cumul}[1]{\left\langle\!\left\langle #1 \right\rangle\!\right\rangle}

\newcommand{\sample}[1]{\left\{ #1 \right\}}
\newcommand{\N}{N_{\rm evts}}
\newcommand{\Np}{N'}

\begin{document}

\preprint{Saclay-T03/098,TIFR/TH/03-15}

\title{Analysis of anisotropic flow with Lee-Yang zeroes}

\author{R. S. Bhalerao}
\email{bhalerao@theory.tifr.res.in}
\affiliation{Department of Theoretical Physics, Tata Institute
of Fundamental Research, Homi Bhabha Road, Colaba, Mumbai 400 005,
India}

\author{N. Borghini}
\email{borghini@spht.saclay.cea.fr}

\author{J.-Y. Ollitrault}
\email{Ollitrault@cea.fr}
\altaffiliation[also at ]{L.P.N.H.E., Universit{\'e} Pierre et Marie Curie,
4 place Jussieu, F-75252 Paris cedex 05, France}
\affiliation{Service de Physique Th{\'e}orique, CEA-Saclay,
F-91191 Gif-sur-Yvette cedex, France}

\date{\today}

\begin{abstract}
We present a new method to extract anisotropic flow in heavy ion
collisions from the genuine correlation among a large number of 
particles. Anisotropic flow is obtained from the zeroes in the complex 
plane of a generating function of azimuthal correlations, in close 
analogy with the theory of phase transitions by Lee and Yang.
Flow is first estimated globally, i.e., averaged over the
phase space covered by the detector, and then differentially, as
a function of transverse momentum and rapidity for identified
particles. 
The corresponding estimates are less biased by nonflow
correlations than with any other method. 
The practical implementation of the method is rather straightforward.
Furthermore, it automatically takes into account most corrections
due to azimuthal anisotropies in the detector acceptance. 
The main limitation of the method is statistical errors,
which can be significantly larger than with the ``standard'' method of
flow analysis if the flow and/or the event multiplicities are too small.
In practice, we expect this to be the most accurate method
to analyze directed and elliptic flow in fixed-target heavy-ion 
collisions between 100~MeV and 10~GeV per nucleon (at the Darmstadt
SIS synchrotron and the Brookhaven Alternating Gradient Synchrotron),
and elliptic flow at ultrarelativistic energies (at the Brookhaven
Relativistic Heavy Ion Collider, and the forthcoming Large Hadron
Collider at CERN).
\end{abstract}


\maketitle

\section{Introduction}

Study of anisotropic flow of particles~\cite{Reviews} produced in
relativistic heavy-ion collisions has emerged as an important tool to
probe the early history, especially the thermalization, of the dense
fireball produced in these collisions.
Anisotropic flow means that the azimuthal distribution of particles
produced in non-central collisions, measured with respect to the
direction of impact parameter, is not flat.
This is characterized by the Fourier harmonics~\cite{Voloshin:1996mz}:
\begin{equation}
\label{defivn}
v_n\equiv \mean{\cos n(\phi-\Phi_R)},
\end{equation}
where $\phi$ denotes the azimuthal angle of an outgoing particle,
$\Phi_R$ is the azimuthal angle of the impact parameter
($\Phi_R$ is also called the orientation of the reaction plane),
$n$ is a positive integer, and angular brackets denote an
average over many particles belonging to some phase-space region,
and over many events.
Throughout this paper, we assume that colliding nuclei
are spherical and parity is conserved: then, the system
is symmetric with respect to the reaction plane and
$\mean{\sin n(\phi-\Phi_R)}$ vanishes for all $n$.
The two lowest harmonics $v_1$ and $v_2$ are named
directed and elliptic flows.
They have been studied extensively over the last several years, with
reference to the wealth of data produced at GANIL~\cite{Cussol:2001df}, 
at the Darmstadt SIS synchrotron~\cite{FOPI,Andronic:2000cx,Andronic:2001sw,%
  Taranenko:1999yh}, 
the Brookhaven Alternating Gradient Synchrotron 
(AGS)~\cite{Jain:1995cm,Barrette:xr,E877,Pinkenburg:1999ya,Liu:2000am,E895,%
  Chung:2001qr}, 
and the Super Proton Synchrotron (SPS) at CERN~\cite{CERES,Agakichiev:2003gg,%
  NA49,NA49new,WA98},
as well as the new and upcoming data from the Relativistic Heavy Ion Collider 
(RHIC) at 
Brookhaven~\cite{PHENIX,PHENIX200,PHOBOS,STAR2p,Adler:2002pu,Adler:2002ct}.

Since the reference direction $\Phi_R$ is not known experimentally,
measuring $v_n$ is a difficult task.
An alternate idea is to extract it from the
interparticle correlations which arise indirectly
due to the correlation of each particle with the reaction plane.
However, standard methods of correlating a particle with an estimate of the 
reaction plane~\cite{Danielewicz:hn,Ollitrault:1997di,Poskanzer:1998yz},
or of correlating two particles with each other~\cite{Wang:1991qh} were
shown to be inadequate at ultrarelativistic SPS
energies~\cite{Dinh:1999mn} due to the smallness of the flow,
and the comparatively large magnitude of
``nonflow'' correlations~\cite{Poskanzer:1998yz,Ollitrault:dy}
due to transverse momentum conservation, resonance decays, etc.,
which are usually neglected in these methods.
At the higher RHIC energies, it was argued that nonflow correlations
due to minijets could even dominate the measured
correlations~\cite{Kovchegov:2002nf}.

These shortcomings of conventional methods 
motivated the development of new methods based on a cumulant
expansion of multiparticle
correlations~\cite{Borghini:2000sa,Borghini:2001vi,Borghini:2002vp}.
The cumulant of the $k$-particle correlation, where $k$ is
a positive integer, isolates the genuine $k$-particle correlation
by subtracting the contribution of lower-order correlations.
Nonflow correlations, which generally involve only a small
number of particles (typically, two or three in the case
of resonance decays, possibly more in the case of minijets)
contribute little to the cumulant if $k$ is large enough.
Note that this is also true for correlations from global
momentum conservation, although the latter involves all
particles~\cite{Borghini:2003ur}.
On the other hand, anisotropic flow is a genuine collective
effect, in the following specific sense:
in a given event, all azimuthal angles are correlated with the
reaction plane azimuth $\Phi_R$, which varies randomly from
one event to the other.
In the laboratory frame, where $\Phi_R$ is unknown, this results
in azimuthal correlations between all particles, or at least a
significant fraction of the particles. 
Therefore, unlike nonflow correlations, anisotropic flow contributes to 
cumulants of all orders $k$.
Neglecting the contribution of nonflow correlations to the cumulant, 
one thus obtains an estimate of $v_n$, which was denoted by
$v_n\{k\}$ in Ref.~\cite{Borghini:2001vi}, and becomes more and
more reliable as $k$ increases.

Cumulants of four-particle correlations were first used to measure
elliptic flow at RHIC~\cite{Adler:2002pu}.
However, it was argued that experimental results
could still be explained by nonflow correlations
alone~\cite{Kovchegov:2002cd} at this order.
Higher-order cumulants, of up to 8 particles, were constructed
at SPS~\cite{NA49new}, and provide the first quantitative
evidence for anisotropic flow at ultrarelativistic energies.
In practice, the cumulant of 8-particle correlations is obtained
by evaluating numerically the 8$^{\rm th}$ derivative of
a generating function. This is rather tedious and numerically
hazardous.

In this paper, we propose to study directly the large-order
behavior of the cumulant expansion, rather than computing
explicitly cumulants at a given order.
Correlating a large number of particles
is the most natural way of studying genuine collective motion
in the system.
Furthermore, finding the large-order behavior turns
out to be simpler in practice than working at a given, finite order.
As we shall see, the large-order behavior is determined by the
location of the zeroes of a generating function in the complex 
plane~\cite{bbo}.
In this respect, our method is in close analogy with the theory
of phase transitions formulated 50 years ago by
Lee and Yang~\cite{Yang:be,Lee:1952ig}:
at the critical point, long-range correlations appear in the
system; as a consequence, the zeroes of the grand
partition function come closer and closer to the real axis
as the size of the system increases. 
A similar phenomenon occurs here, and anisotropic flow appears as formally 
equivalent to a first order phase transition.

The idea of studying anisotropic flow through the correlation
of a large number of particles is not new.
The same idea underlies global event analyses
through a three-dimensional sphericity tensor~\cite{Danielewicz:we},
which led to the first observation of collective flow at
Bevalac~\cite{Gustafsson:ka}.
A similar two-dimensional analysis, restricted to the
transverse plane~\cite{Voloshin:1996mz,Ollitrault:bk},
led to the discovery of flow at the Brookhaven Alternating
Gradient Synchrotron (AGS)~\cite{Barrette:xr}.
With these methods, however, one could not study anisotropic
flow differentially, that is, as a function of transverse momentum
and rapidity for identified particles. This is why they
were soon superseded by the more detailed, although less reliable,
two-particle methods.
One of our important results is that the method presented in this paper
also allows one to analyze differential flow, and its implementation
is rather straightforward.

The paper is organized as follows. In Sec.~\ref{s:recipes} we present
in a self-contained manner our recipes to calculate the ``asymptotic''
integrated and differential flows.
The derivations of these results are given in Sec.~\ref{s:zeroes},
where they are related with the Lee-Yang theory.
The remaining Sections are devoted to detailed discussions of the
various sources of error.
It is now well known that (even when anisotropic flow is absent) ``nonflow'' 
correlations may give a spurious result when the various techniques of flow 
analysis are applied.
In Sec.~\ref{s:sensitivity}, we discuss the magnitude of this
spurious flow, and show that it is significantly smaller than
with any previous method: the main point of this paper is that
the present method is the most efficient way to disentangle
anisotropic flow from other effects.
When anisotropic flow is present in the system,
nonflow correlations produce a systematic error, which is estimated in
Sec.~\ref{s:systematic}.
The effect of fluctuations of flow (due for instance to variations
of impact parameter in a given centrality bin)
on our results is discussed in Sec.~\ref{s:fluctuations}.
Statistical uncertainties on the flow estimates within the method
are computed in Sec.~\ref{s:statistical}.
They are the main practical limitation of our method.
Acceptance corrections due to limited azimuthal coverage of the
detectors are discussed in Sec.~\ref{s:detector}.
Section~\ref{s:summary} is a summary.
Finally, four appendices are devoted to further discussions and
calculations.

\section{Recipes for extracting genuine collective flow}
\label{s:recipes}

In this Section, we show how to obtain an estimate of the flow $v_n$
{}from the genuine correlation among a large number of particles.
The fact that azimuthal angles of outgoing particles are correlated
with the azimuth of the reaction plane $\Phi_R$ allows one to construct
in each event a reference direction, which is called an ``estimate of
the reaction plane'' in the standard Danielewicz-Odyniec
method~\cite{Danielewicz:hn}.
This direction is defined by the flow vector of the event, as recalled
in Sec.~\ref{s:flowvector}.
Note that, unlike other multiparticle methods~\cite{Borghini:2001vi},
the present one is based on the flow vector.
\footnote{In Appendix~\ref{s:Gtilde}, we propose an alternate version of 
the present method, which does not make use of the flow vector, but is more 
time-consuming.}

The flow analysis then proceeds in two successive steps. The first
one is to estimate how the flow vector is correlated
with the true reaction plane.
More precisely, we estimate the mean projection of the
flow vector on the true reaction plane.
This quantity is a weighted sum of the individual flows $v_n$ of all
particles over phase space, which we call ``integrated flow'' and denote by
$V_n$.
The method used to estimate $V_n$ experimentally is described
in Sec.~\ref{s:recipeint}.
This first step is the equivalent of the subevent method in the
standard flow analysis, from which one estimates the
event-plane resolution.
Here however, as well as in the cumulant method based on the flow
vector~\cite{Borghini:2000sa}, subevents are not needed.

The second step in the analysis is to use this reference integrated flow to
analyze ``differential flow,'' i.e., flow in a restricted phase-space window
(e.g., as a function of transverse momentum and rapidity for a given
particle type), which is the goal of the flow analysis.
When analyzing a $p$-th harmonic of the differential flow ($v_p$), the flow 
vector can be chosen in the same harmonic $p$, or in a lower harmonic $n$ 
whose $p$ is a multiple.
One chooses the option which yields the best accuracy. 
\footnote{Note also that the only way to determine the 
sign of $v_p$, with $p\ge 2$, is to use a flow vector in a lower 
harmonic $n<p$. Thus, at SPS the sign of elliptic flow $v_2$ is 
determined using the reference from directed flow $v_1$, while 
its magnitude is determined more accurately using the flow 
vector in the second harmonic~\cite{NA49new}.}
For instance, differential elliptic flow $v_2$ can be analyzed using as a 
reference either the integrated elliptic flow (at ultrarelativistic 
energies~\cite{CERES,NA49,NA49new,WA98,PHENIX,PHENIX200,PHOBOS,STAR2p,%
  Adler:2002pu}) or the integrated directed flow (at lower
energies~\cite{Demoulins:ac,Andronic:2000cx,Chung:2001qr}).

The method to extract differential flow is described in
Sec.~\ref{s:recipedif}.
Following the notations of Refs.~\cite{Borghini:2000sa,Borghini:2001vi},
differential flow in the Fourier harmonic $p$ will be denoted by $v'_p$. 
The estimates of $V_n$ and $v'_p$ obtained
with the present method will be denoted by
$V_n\{\infty\}$ and $v'_p\{\infty\}$, respectively,
where the symbol $\infty$ means that it corresponds to the large-order
behavior of the cumulant expansion, as will be shown in
Sec.~\ref{s:zeroes}.

Section~\ref{s:recipedetector} discusses briefly acceptance issues arising
when the detector does not have full azimuthal coverage.
Section~\ref{s:recipestat} shortly deals with statistical errors,
which are the main limitation of our method.
Both issues are discussed in more detail in Sec.~\ref{s:detector} and
Sec.~\ref{s:statistical}, respectively.

\subsection{The flow vector}
\label{s:flowvector}

The first step of the flow analysis is to evaluate, for each
event, the flow vector of the event.
It is a two-dimensional vector ${\bf Q}=(Q_x,Q_y)$ defined as
\begin{eqnarray}
\label{flowvector}
Q_x&=&\sum_{j=1}^M w_j\,\cos(n\phi_j)\cr
Q_y&=&\sum_{j=1}^M w_j\,\sin(n\phi_j),
\end{eqnarray}
where $n$ is the Fourier harmonic under study ($n=1$ for directed
flow, $n=2$ for elliptic flow),
the sum runs over all detected particles, 
$M$ is the observed multiplicity of the event,
$\phi_j$ are the azimuthal angles of the particles measured with
respect to a fixed direction in the laboratory.

The coefficients $w_j$ in Eq.~(\ref{flowvector}) are weights depending
on transverse momentum, particle mass and rapidity.
The best choice of $w_j$ is that leading to the smallest statistical error, 
by maximizing the flow signal. 
The weight should ideally be proportional to the flow itself,
$w_j(p_T,y)\propto v_n(p_T,y)$~\cite{Borghini:2000sa}.
Otherwise, reasonable choices are $w\propto y-y_{\rm cm}$, i.e., the rapidity
in the center-of-mass frame, for directed flow, and $w=p_T$ for elliptic flow.

When comparing events with different multiplicities $M$,
one may in addition have weights depending on $M$.
Weights proportional to $1/\sqrt{M}$ were used in Ref.~\cite{Borghini:2000sa},
and proportional to $1/M$ in Ref.~\cite{Borghini:2001vi}, to minimize the
effect of multiplicity fluctuations.
This complication is unnecessary here.
This issue is discussed in Appendix~\ref{s:equivalence}.

The discussions in this paper will be illustrated by explicit
numerical and analytical examples, in which we choose unit
weights $w_j=1$, for the sake of simplicity.

The flow vector was first introduced in Ref.~\cite{Danielewicz:hn}.
The azimuthal angle $n\Phi_n$ of ${\bf Q}$ is conventionally used in the
analysis of anisotropic flow~\cite{Poskanzer:1998yz} in order to
estimate the orientation of the reaction plane of the event.

Our method uses the projection of ${\bf Q}$ on a fixed,
arbitrary direction making an angle $n\theta$ with respect
to the $x$-axis. We denote this projection by
$Q^\theta$: 
\begin{eqnarray}
\label{defqtheta}
Q^\theta &\equiv& Q_x\cos(n\theta)+Q_y\sin(n\theta)\cr
&=& \sum_{j=1}^M w_j\,\cos(n(\phi_j-\theta)).
\end{eqnarray}
As we shall see, the whole flow analysis can in principle
be performed using this projection of the flow vector
on a fixed direction.
One thus obtains an estimate of the integrated flow
$V_n^\theta\{\infty\}$ (Sec.~\ref{s:recipeint}),
which is then used as a reference to derive an estimate of the differential 
flow $v^{\prime\theta}_{mn}\{\infty\}$ (Sec.~\ref{s:recipedif}).

In practice, however, one should perform the analysis for several
equally spaced values of $\theta$, typically $\theta=(k/p)(\pi/n)$
with $k=0,\ldots,p-1$ and $p=4$ or 5.
This gives several values of $V_n^\theta\{\infty\}$ and
$v^{\prime\theta}_{mn}\{\infty\}$, which are then averaged over $\theta$.
This yields our final estimates of integrated flow, $V_n\{\infty\}$, 
and differential flow, $v^{\prime}_{mn}\{\infty\}$.
As we shall see in Sec.~\ref{s:statistical}, they have
smaller statistical error bars than each individual
$V_n^\theta\{\infty\}$ and $v^{\prime\theta}_{mn}\{\infty\}$.

\subsection{Integrated flow}
\label{s:recipeint}

Integrated flow is defined as the average value of the
flow vector projected on the unit vector with angle
$n\Phi_R$ (for $n=1$, this is the reaction plane):
\begin{eqnarray}
\label{defVn}
V_n
&\equiv& \mean{Q_x\,\cos(n\Phi_R)+Q_y\,\sin(n\Phi_R)}\cr
&=& \mean{Q^{\Phi_R}},
\end{eqnarray}
where we have used the notation introduced in Eq.~(\ref{defqtheta}).
This quantity was denoted by $\bar Q$ in Ref.~\cite{Ollitrault:1997di},
and by $\mean{\bf Q}$ in Ref.~\cite{Borghini:2000sa}.
Note that unlike the flow coefficients $v_n$, which are dimensionless,
our integrated flow involves the weights present in the flow
vector (\ref{flowvector}), which can have a dimension.
Since the flow vector definition involves a sum over all particles, 
the integrated flow scales like the multiplicity $M$. 
With unit weights,
\begin{equation}
\label{Mvn}
V_n=Mv_n,
\end{equation}
where $v_n$ in the right-hand side (rhs) is to be understood as an
average over the phase space covered by the
detector acceptance, and we have neglected fluctuations
of the multiplicity $M$ for simplicity. 

Let us now explain how an estimate of $V_n$ can be obtained in a real
experiment.
For a given value of $\theta$, we first define the following
generating function, which depends on an arbitrary complex variable $z$:
\begin{equation}
\label{defGint}
G^\theta(z)\equiv \sample{e^{z Q^\theta}},
\end{equation}
where curly brackets denote an average over a large number of
events, $\N$, with (approximately) the same centrality.
This generating function has the symmetry properties
\begin{equation}
\label{symmetry1}
G^{\theta+\pi/n}(z)=G^\theta(-z)
\end{equation}
(following from $Q^{\theta+\pi/n}=-Q^\theta$), and
\begin{equation}
\label{symmetry2}
[G^\theta(z)]^*=G^\theta(z^*),
\end{equation}
where the star denotes complex conjugation.
The second identity simply expresses that $G^\theta(z)$ is real for
real $z$.
An alternative choice of the generating function is presented 
in Appendix~\ref{s:Gtilde}. 

In order to obtain the integrated flow, one must evaluate $G^\theta(z)$ 
for a large number of values of $z$ on the upper half of the imaginary 
axis (that is, $z=i r$ with $r$ real and positive). 
One must then 
take the modulus [recall that $G^\theta(z)$ is a complex number], 
$\left|G^\theta(i r)\right|$, and plot it as a function of $r$.
On the imaginary axis, the symmetry properties,
Eqs.~(\ref{symmetry1}) and (\ref{symmetry2}) translate into
the following relations:
$G^{\theta+\pi/n}(ir)=G^\theta(-ir)=[G^\theta(ir)]^*$.
Taking the modulus, this yields
$|G^{\theta+\pi/n}(ir)|=|G^\theta(-ir)|=|G^\theta(ir)|$.
This identity shows that $\theta$ and $\theta+\pi/n$ are equivalent, 
so that one can restrict $\theta$ to the interval $[0,\pi/n]$. 
It also shows that $z=ir$ and $z=-i r$ are equivalent, which is the 
reason why we restrict ourselves to positive $r$ values.

\begin{center}
\begin{figure}[ht!]
\centerline{\includegraphics*[width=\linewidth]{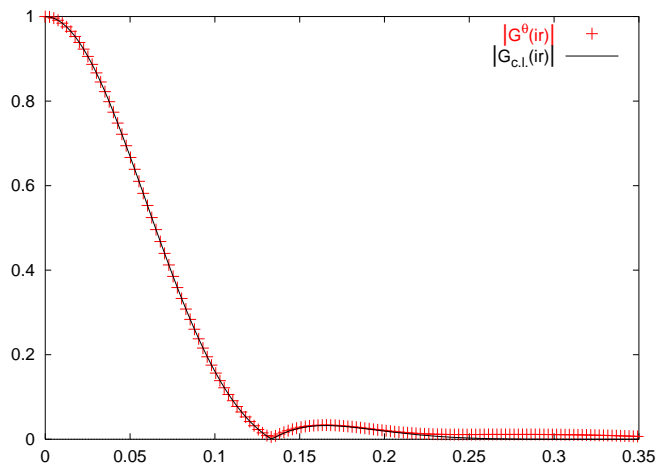}}
\caption{Variation of $\left|G^\theta(ir)\right|$ with $r$,
for $\N=20000$
simulated events, with $M=300$ particles per event and a mean elliptic
flow $v_2=6\%$, to match the typical value in a mid-central Au+Au collision at
$\sqrt{s_{_{\rm NN}}} = 130$~GeV, as analyzed by the STAR Collaboration
at RHIC~\cite{STAR2p}.
The crosses are the values of $|G^\theta(ir)|$ for $\theta=0$.
The solid line displays the expected value $|G_{\rm c.l.}(ir)|$
[defined by Eq.~(\ref{meanGflow})].}
\label{fig:Gvsr}
\end{figure}
\end{center}

To illustrate the variation of $\left|G^\theta(ir)\right|$ as a function 
of $r$, we have computed it for simulated data.
The data set contained $\N=20000$ events with multiplicity $M=300$.
Each event consisted of 10 ``differential bins'' with equally spaced
elliptic flow values ranging from 4.2 to 7.8\%, resulting in an average
$v_2=6\%$, and with a higher flow harmonic $v_4=3\%$ for all particles. 
The flow vector, Eq.~(\ref{flowvector}), was constructed using
unit weights $w_j=1$ and $n=2$. We assumed that the detector
acceptance had perfect azimuthal symmetry. The function
$\left|G^\theta(ir)\right|$ is shown in Fig.~\ref{fig:Gvsr} for $\theta=0$.
Due to rotational symmetry (for a perfect detector), the behavior would be 
similar for another value of $\theta$, up to statistical fluctuations.
The function $\left|G^\theta(ir)\right|$ starts at a value
of $1$ for $r=0$, and it then decreases and oscillates.

Our estimate of integrated flow is directly related to the first minimum
of $\left|G^\theta(ir)\right|$.
Let us denote by $r_0^\theta$ the value of $r$ corresponding to the
first minimum.
The corresponding estimate of integrated flow is defined as
\begin{equation}
\label{flowestimate0}
V_n^\theta\{\infty\}\equiv\frac{j_{01}}{r_0^\theta},
\end{equation}
where $j_{01}\simeq 2.405$ is the first root of the Bessel function $J_0(x)$.
This result will be justified in Sec.~\ref{s:zeroes}.
As one can see in Fig.~\ref{fig:Gvsr}, the variation of
$\left|G^\theta(ir)\right|$ near its minimum is quite steep.
Therefore, from the numerical point of view, one should rather
determine the minimum of the square modulus
$\left|G^\theta(ir)\right|^2$.
\footnote{The following procedure can be used. One first computes
\[
f_N^\theta\equiv \left|G^\theta\!\left(\frac{ij_{01}}{V_{\rm max}
-N\epsilon}\right)\right|^2,
\]
where $V_{\rm max}$ is some {\em a priori} upper bound on the expected
value of the integrated flow $V_n$, $\epsilon$ is a small increment,
and $N$ is an integer varying between $0$ and $V_{\rm max}/\epsilon-1$.
Next, denote by $N_0$ the first value of $N$ for which
$f_{N+1}^\theta>f_N^\theta$, so that $f_{N_0}^\theta<f_{N_0-1}^\theta$
and $f_{N_0}^\theta<f_{N_0+1}^\theta$.
Then, the integrated flow is approximately given by
\[
V_n^\theta\{\infty\} \simeq V_{\rm max}-N_0\epsilon+
\frac{\epsilon\,(f_{N_0+1}^\theta-f_{N_0-1}^\theta)}
{2(f_{N_0-1}^\theta-2 f_{N_0}^\theta+f_{N_0+1}^\theta)}.
\] }

As will be shown in Sec.~\ref{s:zeroes}, the value of 
$|G^\theta(i r)|$ at its minimum would be zero in the limit 
of an infinite number of events. 
One can check experimentally that it is compatible with zero 
within expected statistical fluctuations. 
Anticipating on Sec.~\ref{s:fluctsample} [see the discussion 
following Eq.~(\ref{Absdelta}], the following 
inequality should hold with 95\% confidence level:
\begin{equation}
\label{zerocheck}
|G^\theta(ir_0^\theta)|<\frac{2}{\sqrt{\N}},
\end{equation}
where $\N$ is the number of events used in the analysis.

Eventually, the final estimate $V_n\{\infty\}$ is obtained by averaging
$V_n^\theta\{\infty\}$ over $\theta$:
\begin{equation}
\label{av_over_theta}
V_n\{\infty\}\equiv 
\frac{1}{p}\sum_{k=0}^{p-1}V_n^{k\pi/pn}\{\infty\},
\end{equation}
with $p$ typically 4 or 5.

\begin{center}
\begin{figure}[ht!]
\centerline{\includegraphics*[width=\linewidth]{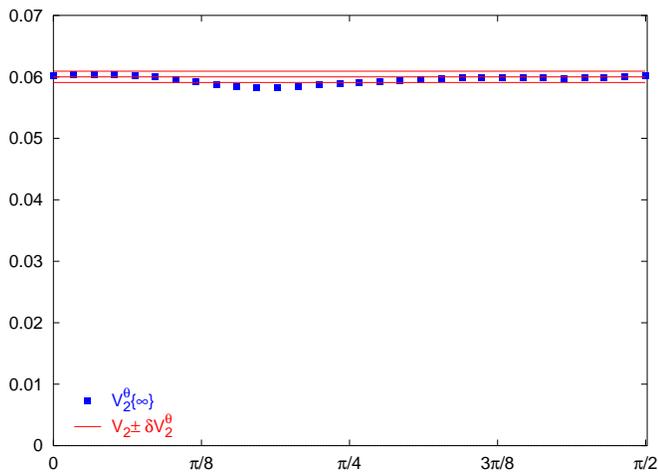}}
\caption{Reconstruction of integrated flow.
The simulated data and the definition of the flow vector are the same
as in Fig.~\ref{fig:Gvsr}.
The plot shows $V_2^\theta\{\infty\}/M$ as a function of $\theta$.
The lines display the expected value with statistical error bars calculated
as explained in Sec.~\ref{s:statint}.}
\label{fig:v2theta}
\end{figure}
\end{center}

This procedure was applied to the simulated data used for Fig.~\ref{fig:Gvsr}.
As stated above, we used constant unit weights $w_j=1$ in 
Eq.~(\ref{flowvector}), so that Eq.~(\ref{Mvn}) holds.
Estimates $V_2^\theta\{\infty\}$ were then computed for many values of
$\theta$ (for the sake of illustration, we did not restrict ourselves
to 4 or 5 values, as would have been enough, see Sec.~\ref{s:statint}).
Since we assumed that the detector is perfect, $V_2^\theta\{\infty\}$ should 
be independent of $\theta$, up to statistical fluctuations. 
The results of the analysis are displayed in Fig.~\ref{fig:v2theta}, where
$V_2^\theta\{\infty\}/M$ is plotted as a function of $\theta$.
The $\theta$-dependence is smooth and has a small amplitude. 
The estimates for $\theta=0$ and 
$\theta=\pi/2$ coincide, as expected from the ($\pi/n$)-periodicity. 

The salient feature in Fig.~\ref{fig:v2theta} is the accuracy of the 
reconstruction of the input. 
As will be shown in Sec.~\ref{s:statint},
the expected statistical uncertainty
on each individual estimate $V_2^\theta\{\infty\}/M$ is $0.092\%$.
One sees in the figure that most values of $V_2^\theta\{\infty\}/M$
are within error bars.
The expected statistical error bar on the average over $\theta$,
$V_2\{\infty\}/M$, is $0.051\%$ (see Sec.~\ref{s:statint}),
i.e., smaller than the statistical error on each
$V_2^\theta\{\infty\}$ by almost a factor of 2.
We obtain $V_2\{\infty\}/M=5.95\%$, in perfect agreement with the
input value $v_2=6\%$.

Finally, let us mention that, like every other method of flow analysis, the 
present one cannot determine the sign of integrated flow: the estimate
given by Eq.~(\ref{flowestimate0}), and its average over $\theta$, 
are positive by definition, while 
the integrated flow defined by Eq.~(\ref{defVn}) can be negative. 
The reason is that our procedure is unchanged if the sign of 
the flow vector is changed in all events [this amounts to changing
$z$ into $-z$ in Eq.~(\ref{defGint})], while this transformation 
changes the sign of the integrated flow $V_n$. 
Strictly speaking, $V_n\{\infty\}$ should therefore be considered 
an estimate of $|V_n|$, rather than $V_n$. 
The sign must be determined independently, or assumed. 

\subsection{Differential flow}
\label{s:recipedif}

Using an estimate of integrated flow in the Fourier harmonic $n$, one
can analyze differential flow in harmonics which are multiples of $n$,
i.e., $mn$, where $m$ is an arbitrary integer.
Following the notations of Ref.~\cite{Borghini:2001vi}, we denote by
$v'_{mn}$ the differential flow corresponding to a given phase-space
window in this harmonic.
We call ``proton'' any particle belonging to the phase-space window under
study, and denote by $\psi$ its azimuthal angle.

For a given value of the angle $\theta$, the corresponding estimate of
differential flow $v'_{mn}$ is given by:
\begin{equation}
\label{diffflow}
\frac{v^{\prime\theta}_{mn}\{\infty\}}{V^\theta_n\{\infty\}}
\equiv
\frac{J_1(j_{01})}{J_m(j_{01})}\,
{\rm Re}\!\left(
\frac{\sample{\cos[mn(\psi-\theta)]\,e^{i r_0^\theta Q^\theta}}}
{i^{m-1}\sample{Q^\theta e^{i r_0^\theta Q^\theta}}}\right),
\end{equation}
where ${\rm Re}$ denotes the real part,
and $r_0^\theta$ has been defined in Sec.~\ref{s:recipeint}.
Two different sample averages appear in the rhs of this equation:
the numerator is an average over all ``protons'' in
all events, while the denominator is an average over {\em events}.
Note that the 
term $\sample{Q^\theta e^{i r_0^\theta Q^\theta}}$ in the denominator
is in fact the derivative of $G^\theta(z)$ with respect to $z$ 
[see Eq.~(\ref{defGint})], evaluated at $z=ir_0^\theta$. 

The numerical coefficient $J_1(j_{01})/J_m(j_{01})$ in
Eq.~(\ref{diffflow}) involves the ratio of two Bessel functions.
It takes the values $1$ for $m=1$ and $j_{01}/2\simeq 1.202$ for $m=2$.
In the case $m=1$ (lowest harmonic), one recovers the estimate
of integrated flow $V^\theta_n\{\infty\}$ by integrating the
corresponding estimate of differential flow
$v^{\prime\theta}_n\{\infty\}$, over all phase space, which amounts to
summing over $\psi$ in the rhs of Eq.~(\ref{diffflow}), with the 
appropriate weighting.

If $\cos$ is replaced by $\sin$ in the numerator of the 
rhs of Eq.~(\ref{diffflow}), the result should be zero 
within statistical errors if parity is conserved. 
This can easily be checked experimentally. 
A non-zero result could be a signature of parity 
violation~\cite{Kharzeev:1998kz,Voloshin:2000xf}. 
This issue will not be discussed further in this paper. 

As in the case of integrated flow, the estimates of differential flow
given by Eq.~(\ref{diffflow}) have a periodicity property, namely
$v^{\prime\,\theta+\pi/n}_{mn}\{\infty\} =
v^{\prime\theta}_{mn}\{\infty\}$.
Indeed, changing $\theta$ into $\theta+\pi/n$ amounts to replacing
the term in brackets by its complex conjugate, which does not affect
the value of the real part.
As above, the estimates $v^{\prime\theta}_{mn}\{\infty\}$ must be 
averaged over $\theta$ in the interval $[0,\pi/n]$, in order to obtain 
an estimate with a reduced statistical uncertainty. 

Please note that there appear ``autocorrelations'' in the numerator of
Eq.~(\ref{diffflow}): one correlates the angle $\psi$ with
$Q^\theta$, which itself involves in general the angle $\psi$
[since the summation in Eq.~(\ref{flowvector}) runs over
all detected particles].
In the standard method of 
flow analysis~\cite{Danielewicz:hn,Poskanzer:1998yz},
autocorrelations are large, so that one has to remove the particle with
angle $\psi$ from the flow vector.
Here, however, autocorrelations do not produce a spurious flow 
by themselves, as will be explained in Sec.~\ref{s:autocorrelations}.
For the lowest harmonic $v'_n$, they lead to a very small correction 
and {\it need not be subtracted\/}.~\footnote{If autocorrelations are 
subtracted, in particular, one no longer recovers exactly the integrated 
flow by integrating the differential flow over phase space.}
For higher harmonics $v'_{2n}$, $v'_{3n}$..., errors due to 
autocorrelations are larger so that one may prefer to subtract
them. However, errors of the same order of magnitude are generally 
expected from nonflow correlations. We come back to this issue 
in Sec.~\ref{s:systematic2}. 

\begin{center}
\begin{figure}[ht!]
\centerline{\includegraphics*[width=\linewidth]{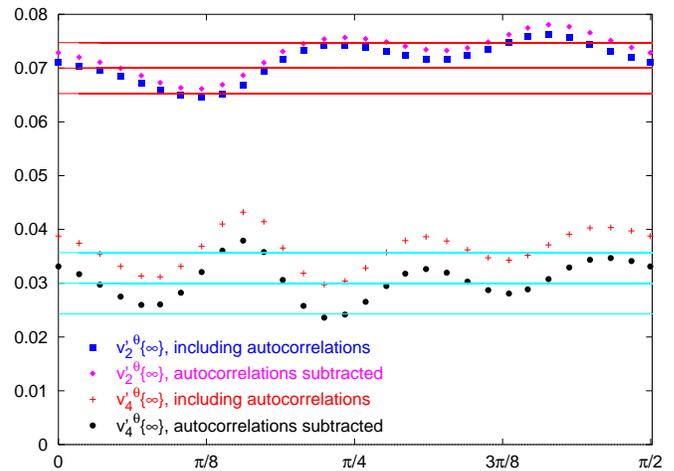}}
\caption{Reconstruction of differential flow.
The simulated data and the definition of the flow vector are the same
as in Figs.~\ref{fig:Gvsr} and \ref{fig:v2theta}.
The plot shows $v_2^{\prime\theta}\{\infty\}$ and
$v_4^{\prime\theta}\{\infty\}$ as a function of $\theta$.
Squares: $v_2^{\prime\theta}\{\infty\}$ with autocorrelations. 
Diamonds: $v_2^{\prime\theta}\{\infty\}$ with autocorrelations removed.
Crosses: $v_4^{\prime\theta}\{\infty\}$ with autocorrelations. 
Circles: $v_4^{\prime\theta}\{\infty\}$ with autocorrelations removed.
The solid lines display the expected value with statistical error
bars calculated as explained in Sec.~\ref{s:statdiff}.}
\label{fig:vdiff}
\end{figure}
\end{center}

Figure~\ref{fig:vdiff} displays the result of the analysis of the same 
Monte-Carlo simulation as in Figs.~\ref{fig:Gvsr} and \ref{fig:v2theta}, 
using a detector with perfect azimuthal symmetry.
We present the differential flow results in ``bin 8,'' corresponding to 
input values of $v'_2=7\%$, $v'_4=3\%$,  
and an average proton number of approximately 30 per event, i.e., 
a total number of protons $\Np\simeq 6\times 10^5$.  
Two different analyses were performed.
The first one followed the procedure presented in this Section.
In a second analysis, we corrected for autocorrelations, subtracting the
contribution of the proton with angle $\psi$ from the flow vector before
multiplying $\cos[mn(\psi-\theta)]$ by $e^{i r_0^\theta Q^\theta}$ in
the numerator of Eq.~(\ref{diffflow}).
One sees in Fig.~\ref{fig:vdiff} that the $v'_2$ result is essentially 
insensitive to autocorrelations.
\footnote{A detailed study of systematic errors, carried out in Appendix 
\ref{s:systerrors}, explains why the flow values with autocorrelations 
subtracted [Eq.~(\ref{systvprime})] are {\it slightly larger\/} 
(instead of {\it much smaller\/} in the standard analysis) than when 
autocorrelations are not taken into account [Eq.~(\ref{vnappsyst})]. 
For the present simulation, $\epsilon$ defined in Eq.~(\ref{abreviations}) 
equals approximately 0.012, which explains the relative difference of 
$2.5\%$ between diamonds and squares in Fig.~\ref{fig:vdiff}.}
We shall see in Sec.~\ref{s:statdiff} that the expected statistical 
errors on $v_2^{\prime\theta}\{\infty\}$ and 
$v_4^{\prime\theta}\{\infty\}$ are 0.47\% and 0.57\%, respectively. 
Values mostly fall within the expected range around the input value, 
except for $v'_4$ when autocorrelations are not subtracted.

After averaging over $\theta$, one obtains 
$v'_2\{\infty\}=7.00\pm 0.26\%$ and $v'_4\{\infty\}=3.60\pm 0.29\%$ 
with autocorrelations and 
$v'_2\{\infty\}=7.20\pm 0.26\%$, $v'_4\{\infty\}=3.03\pm 0.29\%$ 
when autocorrelations are removed, 
where statistical uncertainties have been computed with the help of 
formulas given in Sec.~\ref{s:statdiff}.
For the lower harmonic $v'_2$, results are in good agreement with the 
input value $v'_2=7\%$, whether or not autocorrelations are 
subtracted. For the higher harmonic $v'_4$, a discrepancy with 
the input value $v'_4=3\%$ appears when autocorrelations are
not subtracted. This error will be evaluated analytically in 
Sec.~\ref{s:systematic2}. 

Like other methods of flow analysis, the present procedure 
has a global sign ambiguity, due to the fact that the 
sign of the integrated flow $V_n$ cannot be reconstructed. 
In Eq.~(\ref{diffflow}), both $V_n^\theta\{\infty\}$ and 
$r_0^\theta$ are positive. 
If the true integrated flow $V_n$ is also positive, then
our estimate $v^{\prime\theta}_{mn}\{\infty\}$
has the correct sign. 
If, on the other hand, $V_n$ is negative, then 
$v^{\prime\theta}_{mn}\{\infty\}$ should be multiplied 
by $(-1)^m$. 
[equivalently, one can change the sign of $V_n^\theta\{\infty\}$ 
and $r_0^\theta$ in Eq.~(\ref{diffflow})].

\subsection{Acceptance corrections}
\label{s:recipedetector}

The standard event-plane analysis requires that the
flow vector ${\bf Q}$ has a perfectly isotropic distribution
in azimuth. Since real detectors are not perfect, this requires one
to use various flattening procedures~\cite{Poskanzer:1998yz}.
One of the nice features of the cumulant expansion is that
it isolates physical correlations by subtracting out the contribution
of detector asymmetries~\cite{Borghini:2000sa}.
The same occurs here, so that flattening procedures are not
needed. Detector asymmetries can never produce a spurious
flow by themselves with the method described above.
This holds even if the detector has a very limited
azimuthal coverage.

In Sec.~\ref{s:detector}, we demonstrate that the effects of strong 
azimuthal asymmetries in a detector are twofold. 
First, the estimates $V_2^\theta\{\infty\}$ depend on $\theta$ 
[see Eq.~(\ref{vintdetector})]. 
Second, the average estimate $V_2\{\infty\}$ does not exactly coincide 
with the true value, but differs by a multiplicative coefficient (which 
also controls the $\theta$-dependence of $V_2^\theta\{\infty\}$). 
In most cases, however, this coefficient is so close to unity that 
correcting for it is not even necessary.

\begin{center}
\begin{figure}[ht!]
\centerline{\includegraphics*[width=\linewidth]{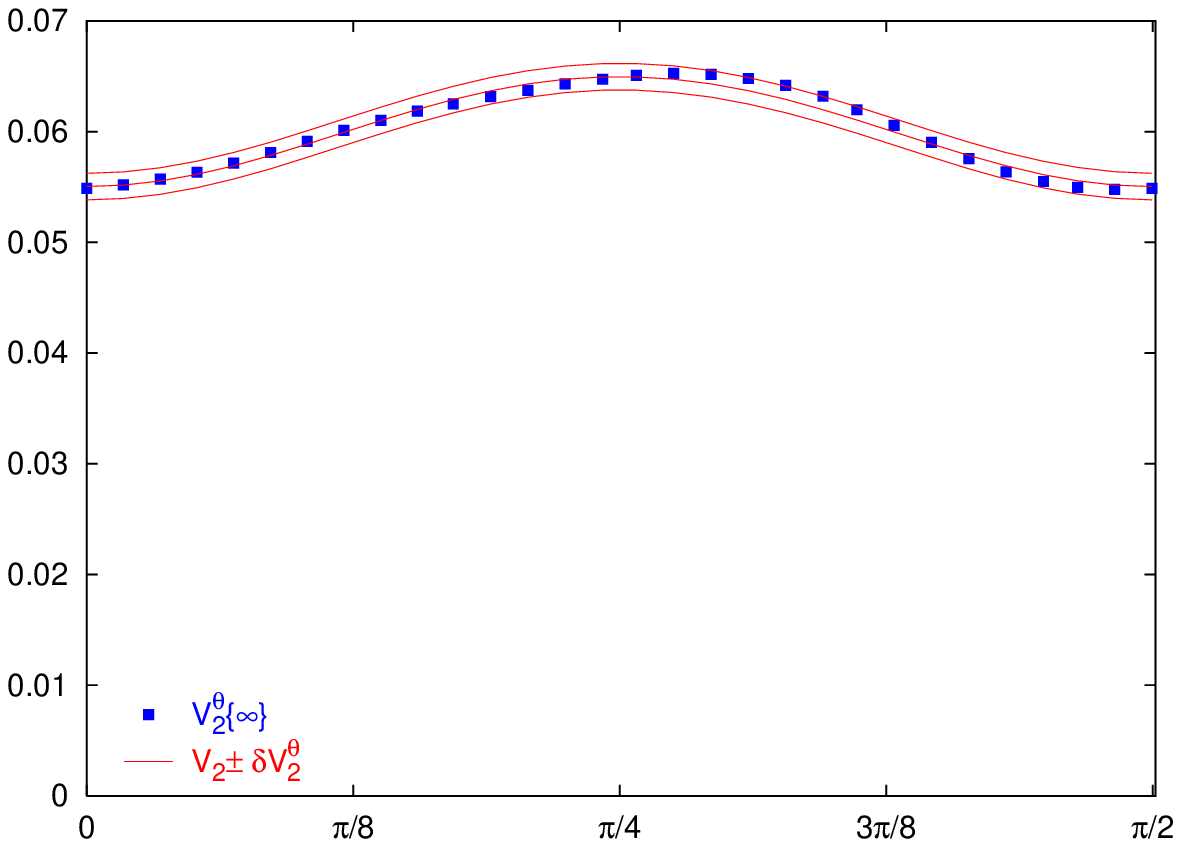}}
\caption{Reconstruction of integrated flow, with a detector which does 
not detect particles with $150^\circ<\phi<210^\circ$.
The simulated data and the definition of the flow vector are the same
as in Fig.~\ref{fig:Gvsr}.
The plot shows $V_2^\theta\{\infty\}/\mean{M}$ as a function of $\theta$.
The lines display the theoretical value taking into account detector 
effects, Eq.~(\ref{vintdetector}), with statistical error bars.}
\label{fig:vacc}
\end{figure}
\end{center}

This is illustrated by Fig.~\ref{fig:vacc}: 
we analyzed a simulated data set with similar statistics as that of 
Figs.~\ref{fig:Gvsr}--\ref{fig:vdiff}, namely 20000 events of 300 
{\em emitted} particles, with an average elliptic flow $v_2=6\%$ (but 
now a vanishing $v_4$). 
We  assumed that the detector had a blind angle of 60~degrees, i.e., that 
one sixth of the azimuthal coverage is missing.
We can clearly see the oscillation of $V_2^\theta\{\infty\}$ as a function 
of $\theta$.
After averaging over $\theta$, however, we find
$V_2\{\infty\}/\mean{M}=6.004\pm 0.066\%$, where $\mean{M}$ denotes 
the mean multiplicity of {\em detected} particles,~\footnote{As a 
consequence of the blind angle, 
the number of detected particles is smaller than the number of 
emitted particles, which was 300 particles for all events in
our simulation, and fluctuates around the mean value $\mean{M}=250$.}
in perfect agreement with the input value $v_2=6\%$.
Note that the statistical error bar is slightly larger than in
Fig.~\ref{fig:v2theta} due to the reduced multiplicity ($\mean{M}=250$
instead of 300). 
With this particular detector configuration, the analytical calculation 
presented in Sec.~\ref{s:detector} shows that the reconstructed 
$V_2\{\infty\}$ should be divided by $1.0068$ in order to take detector 
effects into account [see Eq.~(\ref{vintdetectormean})], a very small 
correction indeed.

Similarly, acceptance inefficiencies have the same two effects on the 
differential flow estimates obtained with the present method. 
Thus, $v^{\prime\theta}_{mn}\{\infty\}$ becomes $\theta$-dependent, 
while $v'_{mn}\{\infty\}$ differs from $v'_{mn}$ by a multiplicative 
factor, see Sec.~\ref{s:detector}. 

Finally, let us emphasize that since the method automatically takes into 
account most detector asymmetries, it is important to take as input in the 
construction of the generating function, Eq.~(\ref{defGint}), the azimuthal 
angles of the particles exactly as they are measured in the laboratory: 
flattening procedures are useless here, and might even bias the analysis.

\subsection{When is the method applicable?}
\label{s:recipestat}

As we shall demonstrate in the following Sections, the main limitation of the 
method arises from statistical errors, while systematic errors inherent to 
the method are under good control. 
In Sec.~\ref{s:statistical}, we shall see that statistical errors
depend on the so-called ``resolution parameter'' $\chi$. 
Referring the reader to Sec.~\ref{s:resolution} for further details on this 
parameter, and in particular on its experimental determination, we simply 
want to recall here that it is roughly given by $\chi\simeq v_n\sqrt{M}$, 
where $M$ is the typical number of detected particles per event, 
and $v_n$ the typical magnitude of the flow. 

Compared to other methods of flow analysis, statistical errors 
depend very strongly on $\chi$. This is because our method involves
the correlation between a large number of particles. 
It is therefore important to use as many particles as possible
in constructing the flow vector, so as to maximize $\chi$. 
In particular, since our 
method is very stable with respect to effects of detector 
acceptance as explained above, one should not do any cuts in phase 
space. Furthermore, since the method is also stable with respect to 
``nonflow'' correlations, one should not worry about possible
``double countings'' due to multiple hits or showering effects
in the detector. 
Most current heavy ion experiments consist of several
detectors of various types (calorimeters, tracking chambers) covering 
various regions of phase space (central rapidity, forward/backward rapidity). 
It is not common in the flow analysis to combine the information 
from different detectors in constructing the flow vector of 
the event. Here, it is important to do so: even a few additional 
particles may result in a significant decrease of the 
statistical error. 
Finally, properly chosen weights can significantly increase 
the value of the resolution parameter $\chi$. 
\footnote{For instance, in the NA49 flow analysis, the values of the 
$p_T$-weighted 
integrated elliptic flow are 20\% larger than the corresponding values 
obtained with unit weights~\cite{NA49new}.}

An over-simplified rule of thumb, which emphasizes the role of 
$\chi$ would be the following:
\begin{itemize}
\item{if $\chi > 1$, the statistical error on the flow is not
    significantly larger than with the standard method 
    (by a factor of at most 2, see Tables~\ref{Tint2}, \ref{Tdiff3}
    and \ref{Tdiff4}), 
    while systematic errors due to nonflow effects are much smaller. 
    Thus, the present method should be used, and statistics will 
    not be a problem;}
\item{if $0.5 < \chi < 1$, the method can be used, but it 
    really is most important to optimize weights, so as to increase $\chi$, 
    possibly by performing two analyses of the same data sets: in the first 
    analysis, adopting (educated) guesses for the weights; and in the second 
    pass, using as weights $w_j$ the differential flow values obtained in
    the first analysis;}
\item{if $\chi < 0.5$, statistical errors are too large, and the
    present method cannot be used; increasing the number of events
    barely helps here; in this case, one should use the cumulant 
    methods of   
    Refs.~\cite{Borghini:2001vi,Borghini:2002vp}, which still apply 
    if the number of events is large enough~\cite{NA49new}.}
\end{itemize}

\section{Collective flow and zeroes of the generating function}
\label{s:zeroes}

Let us now justify the procedure presented in the previous Section.
We begin with a detailed discussion of our hypotheses
(Sec.~\ref{s:hypotheses}).
In Sec.~\ref{s:cumulants}, we define the cumulants of the
distribution of $Q^\theta$. We argue that large-order cumulants
isolate collective effects from other, nonflow correlations,
which generally involve only a few particles.
We then show that large-order cumulants are uniquely
determined by the location of the zeroes of $G^\theta(z)$
in the complex plane of the variable $z$.

Anisotropic flow is a particular case of collective behavior, which
can be analyzed using this procedure.
For this purpose, we compute the generating function $G^\theta(z)$
when anisotropic flow is present (Sec.~\ref{s:Qdistribution}).
This allows us to relate the position of the first zero
of $G^\theta(z)$ to the integrated flow $V_n$.
The analogy of this approach with the Lee-Yang theory of phase
transitions is shown in Sec.~\ref{s:leeyang}.

The discussion of Secs.~\ref{s:cumulants} and \ref{s:Qdistribution}
is generalized to differential flow in Sec.~\ref{s:asympdiff}.
The relation of this approach with the cumulant expansion
of Refs.~\cite{Borghini:2000sa,Borghini:2001vi} is described in
Appendix~\ref{s:equivalence}.

\subsection{Preliminaries}
\label{s:hypotheses}

The derivation of our results relies on the following four hypotheses:
\begin{itemize}
\item{A) the event multiplicity, $M$, is much larger than unity;}
\item{B) for a fixed orientation of the reaction plane $\Phi_R$,
each particle in the system is correlated only to a small number of 
particles, which does not vary strongly with nuclear size and 
impact parameter;}
\item{C) the number of events available, $\N$, is arbitrarily large;}
\item{D) the detector is azimuthally symmetric.}
\end{itemize}

Let us comment briefly on these hypotheses.
With hypothesis C), all sample averages [such as $\sample{\exp(zQ^\theta)}$
in Eq.~(\ref{defGint})] become true statistical averages, which we write
with angular brackets (i.e., $\smean{\exp(z Q^\theta)}$).
Hypothesis A) is verified in practice for heavy-ion collisions at
high energies.
Hypothesis D) is not essential, but simplifies the calculations.
Together with C), it implies for instance
that the generating function, $G^\theta(z)$, is independent of $\theta$.

Hypothesis B) is the crucial assumption, where much physics is
hidden, so let us comment on it in a little more detail.
The idea is that for a fixed geometry (same impact parameter, same 
orientation in space of the colliding nuclei), 
correlations are of the same order of magnitude as if the 
nucleus-nucleus collision was a superposition of independent 
nucleon-nucleon collisions. 
This is a very reasonable assumption at ultrarelativistic energies
\footnote{We could not find a similar argument at lower energies, but 
nevertheless, we feel that the assumption remains valid. 
Anyway, one should be aware that it underlies all methods of collective flow 
analysis.}
if the sample of events used in the analysis have exactly the same
geometry, i.e., exactly the same impact parameter and $\Phi_R$.
Then, nucleon-nucleon collisions occurring in different places
in the transverse plane are uncorrelated, simply by causality:
the transverse size is much larger than the time scale of
the collision due to Lorentz contraction.
Final state interactions may of course induce or (more likely) 
destroy correlations but we assume that they remain short-ranged, 
in the sense that they involve only a few particles, whose number remains
roughly constant as the system size (i.e., the number of participating
nucleons) 
increases.\footnote{Note that we do not exclude the possibility that 
there exist long-range correlations in phase space, for instance in rapidity.}
This hypothesis breaks down in the vicinity of a phase transition, 
where critical fluctuations induce long-range 
correlations~\cite{Stephanov:1999zu}.
This would be even more interesting than anisotropic flow itself. 

Since the overall number of particles is large [hypothesis A)],
hypothesis B) implies that each event can be split in some way into 
a large number of independent ``subevents,'' whose number $N$ 
roughly scales with the system size, i.e., with the multiplicity $M$. 
\footnote{These subevents, which may contain only a few particles, have 
no relation with the ``subevents'' used to determine the event-plane 
resolution in the standard flow-analysis method.}
In practice, this means that if one creates an artificial event by taking
the first subevent from one event, the second subevent from another
event, etc., the resulting ``mixed'' event looks exactly like a real event
(note that we have assumed that all events have exactly the same
reaction plane $\Phi_R$, therefore such mixed events cannot be constructed
experimentally: we use them simply as an image to illustrate our
hypothesis).

If $N$ denotes the number of independent subevents, 
then $Q^\theta$ can be written as the sum of 
$N$ independent contributions: $Q^\theta=\sum_{j=1}^N Q^\theta_j$.
We may write
\begin{equation}
\label{subevents}
\mean{e^{zQ^\theta}|\Phi_R}=
\prod_{j=1}^N \mean{e^{zQ_j^\theta}|\Phi_R},
\end{equation}
where the notation $\smean{\ldots|\Phi_R}$ denotes an average value taken 
over many events having exactly the same reaction plane $\Phi_R$.
The logarithm of this expression scales like the number of independent
subsystems $N$, which itself scales like the multiplicity $M$:
\begin{equation}
\label{shortrange1}
\ln\mean{e^{zQ^\theta}|\Phi_R}\sim
M(a z+b z^2+c z^3\cdots),
\end{equation}
where the coefficients in the expansion are independent of the
system size, and of order unity if the weights in Eq.~(\ref{defqtheta})
are of order unity. 
This equation is the mathematical formulation of hypothesis B),
on which the following discussion relies.
Note that global momentum conservation, although it involves
all particles, effectively behaves as a short-range
correlation~\cite{Borghini:2003ur}, so that it does not invalidate
Eq.~(\ref{shortrange1}).
If the azimuthal distribution is symmetric (i.e., no anisotropic flow,
azimuthally symmetric detector), the left-hand side (lhs) is independent of
$\theta$ by azimuthal symmetry, and Eq.~(\ref{symmetry1}) then implies
that it is an even function of $z$. 
As a consequence, odd terms vanish in the rhs of Eq.~(\ref{shortrange1}). 

\subsection{Cumulants}
\label{s:cumulants}

The cumulants $\cumul{(Q^\theta)^k}$
are defined as the coefficients in the power series expansion
of $\ln G^\theta(z)$~\cite{vanKampen}:
\begin{equation}
\label{defcumulint}
\ln G^\theta(z)=\sum_{k=1}^{+\infty}\frac{z^k}{k!}\cumul{(Q^\theta)^k}.
\end{equation}
If there is no anisotropic flow, outgoing particles are not
correlated to the reaction plane $\Phi_R$. Then,
both sides of Eq.~(\ref{shortrange1}) are independent of $\Phi_R$.
The lhs is simply $\ln G^\theta(z)$ according to the definition, 
Eq.~(\ref{defGint}). 
One concludes that the cumulants
scale linearly with the size of the system, i.e., with the
total multiplicity $M$.

This scaling law breaks down if there is anisotropic flow,
in which case the rhs of Eq.~(\ref{shortrange1}) depends
on $\Phi_R$. It also breaks down if hypothesis B) is not
satisfied, i.e., if there are other collective effects (other than 
global momentum conservation~\cite{Borghini:2003ur}) in the system.
Then, cumulants generally scale with the total multiplicity like
$M^k$. This scaling law is most natural: $Q^\theta$ in
Eq.~(\ref{defqtheta}) is the sum of $M$ terms, so that in
the generating function, Eq.~(\ref{defGint}), $z$ is always
multiplied by $M$ terms. It is therefore natural that
$z^k$ in Eq.~(\ref{defcumulint}) goes with a factor
proportional to $M^k$.
This is the case, in particular, when anisotropic flow is present,
as an explicit calculation in Sec.~\ref{s:Qdistribution} will show.

Therefore, the contribution of collective effects to the cumulants
becomes dominant as $k$ increases.
Physically, the cumulant of order $k$ isolates the genuine
$k$-particle correlation if $k\ll M$:
taking the logarithm in Eq.~(\ref{defcumulint}) effectively
subtracts out the contributions of lower-order correlations,
as shown explicitly in Refs.~\cite{Borghini:2000sa,Borghini:2001vi}.
In order to disentangle collective motion (which involves
by definition a large fraction of the particles) from
lower-order correlations, it is therefore natural to construct
cumulants of order $k$ as large as possible.

The idea of the cumulant expansion proposed
in Refs.~\cite{Borghini:2000sa,Borghini:2001vi} was to construct
explicitly the cumulants at a given order $k$.
Here, we propose to study directly the asymptotic limit when
$k$ goes to infinity, as we now explain.
\footnote{When $k$ becomes as large as the multiplicity $M$, the cumulant 
no longer corresponds to the genuine $k$-particle correlation. 
While the latter cannot be defined for $k$ larger than $M$, the cumulants are 
well defined to all orders.}
The asymptotic behavior of the cumulants
defined by Eq.~(\ref{defcumulint}), when $k$ goes to infinity,
is determined by the
radius of convergence of the power series expansion,
i.e., by the singularities of $\ln G^\theta(z)$ which are closest
to the origin in the complex plane of the variable $z$.
It is obvious from the definition, Eq.~(\ref{defGint}),
that $G^\theta(z)$ has no singularity.
Hence the only singularities of $\ln G^\theta(z)$ are the
zeroes of $G^\theta(z)$.

Let us denote by $z^\theta_0$ the zero of $G^\theta(z)$ which is
closest to the origin in the upper half of the complex plane.
Remember that Eq.~(\ref{symmetry2}) relates the behavior of $G^\theta(z)$
in the lower half of the complex plane to that in the upper half, so that
we only need to study the latter.
The asymptotic behavior of the cumulants for large $k$
is derived in Appendix~\ref{s:largeorders}, Eq.~(\ref{logcauchy}):
\begin{equation}
\label{logcauchybis}
\frac{1}{k!}\cumul{(Q^\theta)^{k}}\sim
-\frac{2}{k}{\rm Re}\!\left(\frac{1}{(z^\theta_0)^k}\right).
\end{equation}
As expected, large-order cumulants depend only on $z_0^\theta$.

Collective effects, if any, are uniquely determined by $z_0^\theta$.
They result in larger correlations, i.e., in higher values
of the cumulants than in the absence of collective motion.
According to Eq.~(\ref{logcauchybis}), this
means that $z_0^\theta$ should be smaller if collective effects
are present. In particular, $z_0^\theta$ will in general come
closer to the origin as the size of the system, $M$, increases,
as we shall see in Sec.~\ref{s:Qdistribution}.
If only short-range correlations are present in the system,
on the other hand, $z_0^\theta$ does not depend on $M$ in the limit of
large $M$. This will be shown by means of an explicit example
in Sec.~\ref{s:sensitivity}.

\subsection{Relation with anisotropic flow}
\label{s:Qdistribution}

We now evaluate the generating function $G^\theta(z)$ in the
presence of anisotropic flow.
We assume that $|z|$ is much smaller than unity (with weights
of order unity), which will be justified later in
Sec.~\ref{s:sensitivity}.
Since all coefficients in the power-series expansion of
Eq.~(\ref{shortrange1}) are of the same order of magnitude,
we can truncate this series for $|z|\ll 1$ by keeping only
the first two terms, which we rewrite in the form
\begin{equation}
\label{centrallimit}
\ln\mean{e^{zQ^\theta}|\Phi_R}\simeq 
\mean{Q^\theta|\Phi_R} z + \frac{\sigma^2 z^2}{4}.
\end{equation}
In this equation, $\smean{Q^\theta|\Phi_R}$ denotes the average value of
$Q^\theta$ for a given $\Phi_R$.
Using the definition of integrated flow, Eq.~(\ref{defVn}),
and symmetry with respect to the reaction plane, one obtains
$\mean{Q_x|\Phi_R}=V_n\cos(n\Phi_R)$ and
$\mean{Q_y|\Phi_R}=V_n\sin(n\Phi_R)$. From the definition
of $Q^\theta$, Eq.~(\ref{defqtheta}), it follows that
\begin{equation}
\label{averageQ}
\mean{Q^\theta|\Phi_R}=V_n\,\cos(n(\Phi_R-\theta)).
\end{equation}
The parameter $\sigma$ in Eq.~(\ref{centrallimit}) is
the standard deviation of $Q^\theta$ for a fixed $\Phi_R$:
\begin{equation}
\label{defsigma2}
\sigma^2\equiv 2\left(\mean{(Q^\theta)^2|\Phi_R}-\mean{Q^\theta|\Phi_R}^2
\right).
\end{equation}
Comparing Eqs.~(\ref{shortrange1}) and (\ref{centrallimit}), $\sigma^2$
scales linearly with the size of the system.
For independent particles, in particular (no correlations, no
anisotropic flow), using Eq.~(\ref{defqtheta}), one obtains
\begin{equation}
\label{sigma2}
\sigma^2\simeq \mean{\sum_{j=1}^M w_j^2}.
\end{equation}

{}From the average value of $e^{z Q^\theta}$ for a fixed $\Phi_R$,
given by Eq.~(\ref{centrallimit}), one deduces
the probability distribution of $Q^\theta$ (for a fixed $\Phi_R$)
by inverse Laplace transform.
It is easily found to be Gaussian: this shows that the approximation
made in keeping only the first two terms in expansion~(\ref{shortrange1})
is the central limit approximation already used
in Refs.~\cite{Voloshin:1996mz,Ollitrault:1997di,Ollitrault:bk}.

With the help of Eq.~(\ref{averageQ}), Eq.~(\ref{centrallimit})
can be rewritten in the form
\begin{equation}
\label{meanGphiR}
\mean{e^{zQ^\theta}|\Phi_R}=
e^{\sigma^2 z^2/4 + V_n\cos(n(\Phi_R-\theta)) z}.
\end{equation}
We now average over $\Phi_R$, so as to obtain $G^\theta(z)$.
Here, we further assume that $\sigma$ in Eq.~(\ref{defsigma2}) is
independent of $\Phi_R$.
The validity of this assumption will be discussed
in Sec.~\ref{s:systematic1}.
The theoretical estimate of $G^\theta(z)$ under this condition
is denoted by $G_{\rm c.l.}(z)$, where the subscript ``c.l.'' refers
to the central limit approximation:
\begin{equation}
\label{meanGflowbis}
G_{\rm c.l.}(z)=e^{\sigma^2 z^2/4}\, I_0(V_n z),
\end{equation}
where $I_0$ is the modified Bessel function of order 0.
The generating function is independent of $\theta$, as
expected from azimuthal symmetry. It is even 
[a natural consequence of Eq.~(\ref{symmetry1}) and azimuthal
symmetry], so that cumulants of odd order vanish.
It is also an even function of $V_n$, so that the sign of $V_n$
cannot be determined from the generating function. From now on, 
we assume that it is positive. 

Anisotropic flow enters the generating function through the 
combination $V_nz$. Therefore, cumulants of even orders $2k$
are of order $(V_n)^{2k}$, i.e., they scale with $M$ like $M^{2k}$
as expected from the general discussion in
Sec.~\ref{s:cumulants}. From the cumulant of order $2k$,
one can obtain an estimate of the flow $V^\theta_n\{2k\}$
in the following way: take the logarithm of Eq.~(\ref{meanGflowbis}),
expand to order $2k$, and identify the coefficient with the
corresponding coefficient in Eq.~(\ref{defcumulint}).
This is roughly the procedure proposed in
Refs.~\cite{Borghini:2000sa,Borghini:2001vi} (a more thorough comparison
is performed in Appendix~\ref{s:equivalence}).
Here, we want instead to take directly the large-order limit
$k\to\infty$.
The large-order expansion of $\ln G_{\rm c.l.}(z)$ is given by
Eq.~(\ref{logcauchy}), where $z_0$ denotes the ``first''
zero of $G_{\rm c.l.}(z)$.
All zeroes of $I_0$ lie on the imaginary axis.
On the imaginary axis, we have
\begin{equation}
\label{meanGflow}
G_{\rm c.l.}(i r)=e^{-\sigma^2 r^2/4}\, J_0(V_n r),
\end{equation}
where $J_0$ is the Bessel function of order zero, and $r$ is real.
The first zero of $J_0$ lies at $j_{01}\simeq 2.405$, hence the first
zero of $G_{\rm c.l.}(z)$ is at
\begin{equation}
\label{defz0}
z_0=ir_0=\frac{ij_{01}}{V_n}.
\end{equation}
Since $V_n$ scales like the total
multiplicity $M$, $z_0$ comes closer to the origin
as $M$ increases, as expected from the discussion in
Sec.~\ref{s:cumulants}.
Note that $G_{\rm c.l.}(z)$ has no zero when there
is no anisotropic flow.

Identifying the terms in $z^{2k}$ in the expansions of
$\ln G_{\rm c.l.}(z)$ and $\ln G^\theta(z)$, one obtains for large $k$
\begin{equation}
\label{vn2k}
V_n^\theta\{2k\}^{2k}=(-1)^k (j_{01})^{2k}\,
{\rm Re}\!\left(\frac{1}{(z^\theta_0)^{2k}}\right).
\end{equation}
If the zero of $G^\theta(z)$ lies exactly on the imaginary
axis at $z_0^\theta=ir_0^\theta$, as the zero of $G_{\rm c.l.}(z)$,
then $V_n^\theta\{2k\}$ converges to a limit for large $k$,
which is $V_n^\theta\{\infty\}$ defined by Eq.~(\ref{flowestimate0}).

If, however, $z_0^\theta$ has a (small) non-vanishing real part,
$V_n^\theta\{2k\}$ does not converge for large $k$.
Unfortunately, this is the general case:
experimentally, the available number of events $\N$ is
finite, and the resulting statistical fluctuations of $G^\theta(z)$
push the zeroes of $G^\theta(z)$ slightly off the imaginary axis.
These deviations, however, are physically irrelevant.
This is why we suggested, in Sec.~\ref{s:recipeint}, to find
the first minimum of $\left| G^\theta(ir)\right|$
(denoted by $r_0^\theta$), rather than the first zero of
$G^\theta(z)$ in the complex plane.
Both procedures are of course equivalent when $z_0^\theta$ lies on the
imaginary axis, as it should in an ideal experiment.
We then approximate $z_0^\theta\simeq i r_0^\theta$
in Eq.~(\ref{vn2k}). Then, we obtain a consistent limit
for large $k$, which is again
Eq.~(\ref{flowestimate0}).
However, one should keep in mind that $V_n^\theta\{\infty\}$
is the limit of $V_n^\theta\{2k\}$ for large $k$ only when
the first zero lies exactly on the imaginary axis.

The careful reader will have noted that the theoretical
estimate $G_{\rm c.l.}(ir)$ is real [see Eq.~(\ref{meanGflow})].
One could then argue that the imaginary part of $G^\theta(ir)$
is also irrelevant, and look for the first zero of the real part
of $G^\theta(ir)$, rather than the first minimum of
$|G^\theta(ir)|$.
However, this is incorrect. We shall see in Sec.~\ref{s:detector}
that small inhomogeneities in the detector acceptance slightly
shift the phase of $G^\theta(ir)$. The real part oscillates
and has zeroes, but they are irrelevant.

The following Sections of this paper are essentially
discussions of the errors which occur when one of the hypotheses
listed in Sec.~\ref{s:hypotheses} is violated.
First, we discuss errors resulting from the simplifications
made in Sec.~\ref{s:Qdistribution}. They will be shown
to amount to finite multiplicity corrections [violations
of hypothesis A)]. These errors are of two types:
1. Even if no anisotropic flow
is present in the system, the generating function does have
zeroes, so that the analysis will yield a spurious flow.
Its magnitude is discussed in Sec.~\ref{s:sensitivity}.
2. When anisotropic flow is present,
there are finite multiplicity corrections to the formula
(\ref{flowestimate0}). These are discussed in
Sec.~\ref{s:systematic}.
Next, we discuss a specific violation of hypothesis B),
namely fluctuations of impact parameter in the sample of
events, in Sec.~\ref{s:fluctuations}.
Violations of hypothesis C) result in statistical errors,
which are computed in Sec.~\ref{s:statistical}.
Finally, violations of hypothesis D), i.e., detector effects,
are discussed in Sec.~\ref{s:detector}.

\subsection{Relation with the Lee-Yang theory of phase transitions}
\label{s:leeyang}

Lee and Yang showed in 1952
that phase transitions can be characterized by the distribution of the
zeroes of the grand partition function in the complex plane~\cite{Yang:be}.
The grand partition function is defined as
\begin{equation}
\label{zmu}
{\cal G}(\mu)=\sum_{N=0}^{\infty} Z_N e^{\mu N/kT},
\end{equation}
where $Z_N$ is the canonical partition function for $N$ particles
at temperature $T$ in a finite volume $V$. Both $T$ and
$V$ are fixed.
In order to make the analogy with our formalism
more transparent, we rewrite the above grand partition
function in the form
\begin{equation}
\label{Gleeyang}
G(z)\equiv \frac{{\cal G}(\mu)}{{\cal G}(\mu_c)}
=\sum_{N=0}^{\infty} P_N e^{zN},
\end{equation}
where
$z\equiv (\mu-\mu_c)/kT$, $P_N\equiv Z_N e^{\mu_cN/kT}/{\cal G}(\mu_c)$,
and $\mu_c$ is a reference value.
Lee and Yang rewrote the grand partition function in a similar
way, but they choose the variable $y=e^z$ (fugacity) instead of $z$.

Physically, the coefficients $P_N$ in Eq.~(\ref{Gleeyang}) represent
the probability of having $N$ particles in the system at chemical
potential $\mu_c$.
The grand partition function can therefore be rewritten as
a statistical average with this probability distribution:
\begin{equation}
\label{Gleeyang2}
G(z)=\mean{e^{zN}}.
\end{equation}
The formal analogy with our generating function, Eq.~(\ref{defGint}), is
obvious.

Lee and Yang studied the repartition of the zeroes of $G(z)$ in the plane of 
the complex variable $z$.
\footnote{These zeroes completely characterize the grand partition function 
if there is a hard-core interaction: in this case, there is an upper bound on 
the multiplicity $N$ for a finite volume, so that $G(z)$ is a polynomial of 
the variable $y=e^z$, which is completely determined by its roots and its 
value at the origin $G(0)=1$. However, most of the crucial results of Lee and 
Yang are still valid if this hard-core assumption is relaxed.}
There is no zero on the real axis, since Eq.~(\ref{Gleeyang}) is a sum of 
positive terms.
Lee and Yang first showed that if a phase transition occurs at
$\mu=\mu_c$, the zeroes of $G(z)$ in the complex plane of the
variable $z$ come closer and closer to the origin $z=0$ as the
volume of the system, $V$, increases~\cite{Yang:be}.
If no phase transition occurs, on the other hand, zeroes remain
at a finite distance from the origin.

This approach can easily be extended to the canonical partition
function, when expressed as a function of the variable
$z\equiv 1/(kT_c)-1/(kT)$, where $T_c$ is the critical temperature.
In this case, one usually speaks of Fisher zeroes~\cite{fisher}
rather than Lee-Yang zeroes.

The properties of $G^\theta(z)$ derived in Sec.~\ref{s:cumulants}
also apply to $G(z)$ as defined by Eq.~(\ref{Gleeyang2}),
provided one replaces $Q^\theta$ by the number
of particles, $N$, and the multiplicity $M$
by the volume $V$. More specifically,
it is well known that if a system can be decomposed into
$n$ independent subsystems, the partition function can be factorized
into the product of the individual partition functions of each subsystem:
$G(z)=\prod_{j=1}^n G_j(z)$. 
Then, the zeroes of $G(z)$ for the whole
system are the zeroes of the partition functions $G_j(z)$
for each subsystem. If the subsystems are equivalent,
the position of the zeroes of $G(z)$ does not change as the number of
subsystems increases. In particular, their distance from the
origin does not decrease as the system size increases.
This property can be generalized to systems with short-range 
correlations.

If long-range correlations are present, on the other hand,
large-order cumulants are larger (see the discussion in
Sec.~\ref{s:cumulants}), and the first zero of $G(z)$
comes closer and closer to the origin as the system size increases.
This can be easily understood in the case of a first-order
phase transition, say, a liquid-gas phase transition. At the
critical point $\mu=\mu_c$, for a large system, one can have
any mixture of the (low-density) gas phase and the (high-density)
liquid phase. The probability distribution $P_N$ in
Eq.~(\ref{Gleeyang}), instead of being
sharply peaked around its average value, is widely spread
between two values $N_{\rm min}$ (pure gas) and $N_{\rm max}$
(pure liquid) which both scale like the volume $V$.
Then, the partition function $G(z)$ depends on the volume
$V$ essentially through the combination $zV$, and consequently its
zeroes scale with the volume like $1/V$.
Anisotropic flow produces a similar phenomenon: its contribution
to the generating function
$G^\theta(z)$ is a multiplicative term $I_0(V_n z)$
in Eq.~(\ref{meanGflowbis}), where $V_n$ scales like the total
multiplicity $M$, and zeroes accordingly scale like $1/M$.
Anisotropic flow appears formally equivalent to a first-order
phase transition in this respect.
The important difference with statistical physics 
is that the system size is much smaller in practice.
As a consequence, zeroes never come very close to the
origin, but the physics involved is essentially
the same.

The analogy with Lee-Yang theory goes one step further.
In a second paper~\cite{Lee:1952ig}, Lee and Yang showed
that for a very general class of interactions, the zeroes
of $G(z)$ are located on the imaginary axis (i.e., on the
unit circle for their variable $y=e^z$).
Here, although we do not have a general proof for this result,
the same property holds in most cases.
In particular, we have seen in Sec.~\ref{s:Qdistribution}
that zeroes resulting from anisotropic flow lie on the
imaginary axis [see Eq.~(\ref{meanGflow})].
The main reason is that our generating function $G^\theta(z)$ is
even (for a perfect detector and an infinite number of events).
Now, we know that zeroes come in conjugate pairs
due to Eq.~(\ref{symmetry2}), i.e., if $z_0$ is a zero,
then $z_0^*$ is also a zero. If $z_0$ lies on the imaginary
axis, $z_0^*=-z_0$, which satisfies the parity requirement.
This parity argument does not show by itself that zeroes
should be on the imaginary axis, but it is nevertheless
crucial in the proof by Lee and Yang, and it is interesting
to note that their result remains valid in the cases of 
interest in the present paper.

Finally, we wish to mention that Lee-Yang zeroes were also used in 
high-energy physics in another context, namely in analyzing multiplicity
distributions.
It was found that the generating function of multiplicity
distributions has zeroes that tend to lie on the unit
circle in the complex plane of the fugacity, as Lee-Yang
zeroes~\cite{dewolf}.
However, it was later shown that this behavior merely reflects general, 
well-known features of the multiplicity distribution~\cite{Brooks:1997kd},
so that this approach does not seem to give new insight on
the reaction dynamics.

\subsection{Differential flow}
\label{s:asympdiff}

Let us now justify the recipes given in Sec.~\ref{s:recipedif} to analyze 
differential flow.
We shall proceed in the same way as for integrated flow.
We start with a discussion of our hypotheses and of their 
implications. 
Next, we define the cumulants of the correlations between an individual 
particle, whose differential flow we are interested in, and the flow 
vector $Q^\theta$.
We discuss the order of magnitude of these cumulants in the case when 
only short-range correlations are present in the system, and derive the 
general expression of their large-order behavior. 
Then, we compute their value in the presence of anisotropic flow, in order 
to relate them to the differential flow. 
Finally, we explain why ``autocorrelations'' are negligible in our approach.

\subsubsection{Preliminaries}

Our hypotheses are the same as in Sec.~\ref{s:hypotheses}. 
Here, we want to analyze the flow in a given phase space window.
We denote by $\psi$ the azimuthal of a particle belonging
to the window (which we call a proton for sake of brevity). 
In order to study its flow, we have to 
correlate it to the flow vector $Q^\theta$, or equivalently, to the 
generating function $e^{z Q^\theta}$. In order to 
study the Fourier harmonic $v'_{mn}$, it is natural to 
construct averages over all protons 
such as $\smean{\cos(mn(\psi-\theta))\,e^{zQ^\theta}}$. 
We can first perform this average for a fixed orientation 
of the reaction plane, $\Phi_R$. Hypothesis B) allows 
us to obtain an equation similar to Eq.~(\ref{subevents}):
the contributions of independent subevents factorize. 
Dividing by $\smean{e^{zQ^\theta}|\Phi_R}$, these 
contributions cancel, except for the subevent containing $\psi$. 
One thus obtains an equation similar to Eq.~(\ref{shortrange1})
\begin{equation}
\label{shortrange2}
\frac{\mean{\cos(mn(\psi-\theta))\,e^{zQ^\theta}|\Phi_R}}
{\mean{e^{zQ^\theta}|\Phi_R}} = 
a'+b'z+c'z^2\cdots,
\end{equation}
where the coefficients $a'$, $b'$, $c'$ are independent of the 
system size (since they only involve the subevent to which $\psi$
belongs), and typically of order unity [if weights in 
Eq.~(\ref{defqtheta}) are of order unity].
This is the mathematical formulation of hypothesis B).

\subsubsection{Cumulants}

We first introduce the following generating function
\begin{equation}
\label{defD(z)}
D^\theta_m(z)\equiv
\sample{\cos(mn(\psi-\theta))\,e^{z Q^\theta}}.
\end{equation}
This generating function has the symmetry properties
\begin{equation}
\label{symmetryd1}
D^{\theta+\pi/n}_m(z)=(-1)^m D^\theta_m(-z)
\end{equation}
and
\begin{equation}
\label{symmetryd2}
\left[D^\theta_m(z)\right]^*=D^\theta_m(z^*),
\end{equation}
which are analogous to Eqs.~(\ref{symmetry1}) and (\ref{symmetry2}),
respectively. 
Furthermore, if the number of events is infinite and if the
detector has perfect azimuthal symmetry, $D^\theta_m(z)$
is independent of $\theta$.

The cumulants $d_k$ are defined by
\begin{equation}
\label{defcumuldiff}
\frac{D^\theta_m(z)}{G^\theta(z)}=\sum_{k=0}^{\infty}\frac{z^k}{k!} d_k,
\end{equation}
where $G^\theta(z)$ is defined by Eq.~(\ref{defGint}).
If there is no anisotropic flow, outgoing particles are not correlated
with the orientation of the reaction plane, and the lhs of 
Eq.~(\ref{shortrange2}) is the lhs of Eq.~(\ref{defcumuldiff}). 
This shows that cumulants are independent of
the system size (and typically of order unity) when there
is no anisotropic flow.

When collective motion is present, on the other hand, cumulants
scale like $M^k$, as we shall see below in the case of anisotropic
flow.
Following the same reasoning as for integrated flow, this shows
that the best sensitivity to collective flow is achieved by
constructing large-order cumulants.

The large-order behavior of the cumulants is determined by the singularities
of the generating function in the lhs of Eq.~(\ref{defcumuldiff}).
Since $D^\theta_m(z)$ is analytic in the whole complex plane,
the singularities are here again the zeroes of $G^\theta(z)$.
The approximate expression of the cumulants is derived in 
Appendix~\ref{s:largeorders}.
It is given by Eq.~(\ref{polecauchy}):
\begin{equation}
\label{polecauchybis}
\frac{1}{k!} d_k \sim 
{\rm Re}\!\left(-\frac{2}{(z_0)^{k+1}} \,
  \frac{D^\theta_m(z_0)}{\sample{Q^\theta e^{z_0 Q^\theta}}}\right),
\end{equation}
where $z_0$ is the first zero of $G^\theta(z)$, and
where we have evaluated the derivative of
$G^\theta(z)$ in the denominator, using definition~(\ref{defGint}).

\subsubsection{Relation with anisotropic flow}

Let us now compute $D^\theta_m(z)$ when there is anisotropic flow.
We use Eq.~(\ref{shortrange2}) and keep only the first term 
in the power series expansion:
\begin{equation}
\label{centrallimit2}
\frac{\mean{\cos(mn(\psi-\theta))\,e^{z Q^\theta}|\Phi_R}}
{\mean{e^{z Q^\theta}|\Phi_R}}
\simeq \mean{\cos(mn(\psi-\theta))|\Phi_R}.
\end{equation}
This amounts to assuming that the proton is not correlated with the
other particles for a fixed orientation of the reaction plane, 
and that there is no autocorrelation, i.e., that the proton 
is given zero weight in the definition of the flow vector, 
Eq.~(\ref{flowvector}).

The denominator of the lhs of Eq.~(\ref{centrallimit2}) 
is given by Eq.~(\ref{meanGphiR}),
while the rhs is given by a relation analogous to Eq.~(\ref{averageQ}):
\begin{equation}
\label{average1}
\mean{\cos(mn(\psi-\theta))|\Phi_R}=v'_{mn}\cos(mn(\Phi_R-\theta)).
\end{equation}
We denote by $D_{m\, {\rm c.l.}}(z)$ the value
of $D^\theta_m(z)$ under the above hypotheses.
Using the last two equations and averaging over $\Phi_R$,
one obtains
\begin{equation}
\label{dmcl}
D_{m\,{\rm c.l.}}(z)=
e^{\sigma^2 z^2/4}\,I_m(V_n z)\, v'_{mn}.
\end{equation}
Combining this identity with Eq.~(\ref{meanGflowbis}), we finally obtain
\begin{equation}
\label{cumuldiffflow}
\frac{D_{m\, {\rm c.l.}}(z)}{G_{\rm c.l.}(z)}
=\frac{I_m(V_n z)}{I_0(V_n z)}\,v'_{mn}.
\end{equation}
The cumulants are obtained by expanding this function in powers
of $z$, as in Eq.~(\ref{defcumuldiff}). The only non-vanishing
terms are those of order $z^{2k+m}$, where $k$ is a positive integer.

One can thus obtain an estimate of differential flow from the
cumulant of order $2k+m$ by expanding both Eqs.~(\ref{defcumuldiff})
and (\ref{cumuldiffflow}) and identifying the coefficients of $z^{2k+m}$
in these expansions.
This requires to know the integrated flow $V_n$, which has been
estimated earlier.

The large-order coefficients in the expansion of Eq.~(\ref{cumuldiffflow})
are given by Eq.~(\ref{polecauchy}), where $D^\theta_m(z)$ and 
$G^\theta(z)$ are replaced by $I_m(V_nz)$ and $I_0(V_n z)$, respectively.
The derivative of $I_0(V_nz)$ with respect to $z$ is
$V_n\,I_1(V_n z)$, so that
\begin{equation}
\frac{d_k}{k!}\sim
{\rm Re}\!\left(\frac{-2}{(z_0)^{k+1}} \,
\frac{I_m(V_n z_0)}{V_n\,I_1(V_nz_0)}\right) v'_{mn}.
\end{equation}
Next, we replace $V_n$ by its estimate Eq.~(\ref{flowestimate0}).
The pole then lies at $z_0=ir_0^\theta$.
Inserting this value into the previous equation, and using the relation
$I_m(i r)=i^mJ_m(r)$, we obtain
\begin{equation}
\frac{d_k}{k!}\sim
{\rm Re}\!\left(\frac{-2 i^{m-1}}{(ir^\theta_0)^{k+1}} \,
\frac{J_m(j_{01})}{J_1(j_{01})}\right)
\frac{v'_{mn}}{V^\theta_n\{\infty\}}.
\end{equation}
This is nonvanishing only if $k=2k'+m$, as expected. 

In order to obtain an estimate of $v'_{mn}$, we must
identify this expression with the large-order expansion
of the measured cumulants, given by Eq.~(\ref{polecauchybis}).
As explained above in the case of integrated flow,
the real part of the first zero, $z_0$, is irrelevant,
so we replace $z_0$ in Eq.~(\ref{polecauchybis})
by $ir_0^\theta$.
One thus immediately recovers Eq.~(\ref{diffflow}).
The remarkable result is that the generating function
for differential flow $D^\theta_m(z)$ need only be
evaluated for a single value of $z$.
In this respect, the analysis is simpler than the cumulant method of 
Refs.~\cite{Borghini:2000sa,Borghini:2001vi}, and more reliable numerically.

\subsubsection{Autocorrelations}
\label{s:autocorrelations}

In the standard event-plane method~\cite{Danielewicz:hn},
differential flow is obtained by correlating the particle
of interest with the flow vector of the event, Eq.~(\ref{flowvector}).
It is necessary to first remove the particle from the
definition of the flow vector, otherwise the resulting
autocorrelation would produce a spurious differential flow.

One must be aware that even after autocorrelations have 
been removed, an error of the same order of magnitude may 
still be present due to nonflow correlations:
consider the simple example where particles are emitted in 
collinear pairs.~\footnote{Such collinear particles are expected
from minijets, as will be seen in Sec.~\ref{s:sensitivity}. 
Decay products from high transverse momentum resonances 
will also be almost collinear.} 
When one correlates a particle with the 
flow vector, if the flow vector involves the other particle
in the pair, the resulting correlation will be exactly of the 
same magnitude as an autocorrelation. 
Conversely, methods which are less biased by nonflow 
correlations, such as higher-order cumulants, 
 are also less biased by autocorrelations~\cite{Borghini:2000sa}.

Here, autocorrelations do not produce any spurious flow by 
themselves. 
To show this, we consider for simplicity a particle
which is not correlated with any other particle, whatever the
reaction-plane orientation, so that its differential flow vanishes.
We are going to check that the above procedure yields
the correct result $v'_{mn}=0$.

Let us denote by $Q^{\prime\theta}$ the value of $Q^\theta$
[Eq.~(\ref{defqtheta})] after subtracting the contribution of the
particle under study:
$Q^{\prime\theta}=Q^\theta-w\cos(n(\psi-\theta))$, where $w$
is the weight associated with the particle.
If the particle with angle $\psi$ is not correlated with the
other particles, the average in Eq.~(\ref{defD(z)}) can be
factorized:
\begin{equation}
\label{defDprime}
D^\theta_m(z)\equiv
\mean{\cos(mn(\psi-\theta))\,e^{zw\cos(n(\psi-\theta))}}
\mean{e^{z Q^{\prime\theta}}}.
\end{equation}
The last term in the rhs can be estimated as in
Sec.~\ref{s:Qdistribution}. This gives an equation
similar to Eq.~(\ref{meanGflowbis}):
\begin{equation}
\label{meanGflowprime}
\mean{e^{z Q^{\prime\theta}}}=e^{\sigma^{\prime 2} z^2/4}\, I_0(V_n z).
\end{equation}
The crucial point is that the particle which has been removed
does not flow by hypothesis, so that 
it does not contribute to the integrated flow
in Eq.~(\ref{defVn}). Therefore, $Q^\theta$ and $Q^{\prime\theta}$
yield the same integrated flow value $V_n$, although their statistical
fluctuations differ in general ($\sigma'<\sigma$).

Therefore, $\smean{e^{z Q^{\prime\theta}}}$ vanishes for the
same value of $z$ as $G^\theta(z)$.
Now, our estimate of differential flow, Eq.~(\ref{diffflow}),
involves the generating function $D^\theta_m(z)$ precisely
at the point $z_0$ where $G^\theta(z)$ vanishes.
This shows that no spurious differential flow appears,
although autocorrelations have not been removed.

On the other hand, when there is differential flow, 
autocorrelations produce spurious higher harmonics of the 
flow. This will be discussed in Sec.~\ref{s:systematic2}.

\section{Sensitivity of the method}
\label{s:sensitivity}

Even if there is no flow in the system, the method presented 
in this paper will give a spurious ``flow'' value.
In this Section, we estimate the order of magnitude of this 
spurious value, and show that it is smaller than with any other 
method of flow analysis. We assume that hypotheses C) and D)
are satisfied, i.e., we neglect statistical fluctuations
and detector effects.

Since no spurious flow appears in the central limit approximation, 
as shown in Sec.~\ref{s:Qdistribution}, we need to go beyond
this approximation. 
In this Section, we choose to work out exactly a simple model, 
but the order of magnitude of our results is general.

The model is the following:
we assume that all particles are emitted in jets, each jet containing $q$
collinear particles, where $q$ is not much larger than unity. 
Since the total multiplicity is $M$, the
total number of jets is $M/q$.
The case $q=1$ corresponds to independent particle emission. 
With unit weights, Eq.~(\ref{defqtheta}) becomes
\begin{equation}
Q^\theta = q\sum_{j=1}^{M/q} \cos(n(\phi_j-\theta)),
\end{equation}
where the $\phi_j$ are the azimuthal angles of the jets.

We assume that the jets are mutually independent and
uniformly distributed in azimuth.
Equation~(\ref{defGint}) then gives
\begin{eqnarray}
\label{Gnoflow+jets}
G^\theta(z)=\left[I_0\!\left(qz\right)\right]^{M/q}.
\end{eqnarray}
The generating function is even and independent of $\theta$,
thanks to the azimuthal symmetry.
Note that $\ln G^\theta(z)$ is proportional to $M$, as expected
from the general discussion in Sec.~\ref{s:cumulants}.

As explained in Sec.~\ref{s:Qdistribution}, the estimate of
$V_n$ from the cumulant of order $2 k$, denoted by
$V_n^\theta\{2 k\}$, is obtained by
expanding $\ln G^\theta(z)$ to order $z^{2k}$ and identifying
with the expansion of $\ln G_{\rm c.l.}(z)$ [Eq.~(\ref{meanGflowbis})]
to the same order.
One thus obtains
\begin{equation}
\label{spurious2k}
V_n^\theta\{2 k\}=q\left(\frac{M}{q}\right)^{1/2k}.
\end{equation}
This spurious flow increases with $q$, i.e., with the magnitude 
of nonflow correlations. There is even a spurious flow for 
$q=1$, i.e., for independent particles. This is a consequence 
of ``autocorrelations'' which appear when expanding $G^\theta(z)$ 
in powers of $z$, and are not present in other methods 
of flow analysis. An alternative choice for the generating function, 
which is free from these autocorrelations, is discussed in 
Appendix~\ref{s:Gtilde}. It will be shown that it does not give
better results as soon as $q\ge 2$, i.e., as soon as nonflow 
correlations are present: when this is the case,
Eq.~(\ref{spurious2k}) still gives the order of magnitude 
of the spurious flow.

As expected from the general arguments in Sec.~\ref{s:cumulants},
the spurious flow decreases as the order $k$ of the cumulant
expansion increases.
We now evaluate $V_n^\theta\{\infty\}$ defined by Eq.~(\ref{flowestimate0}).
Although there is no flow in the system, the generating function
(\ref{Gnoflow+jets}) does have zeroes on the imaginary axis.
The first one lies at $z=ir_0^\theta$, with $r_0^\theta=j_{01}/q$.
[Note that it is a multiple zero, because of the power $M/q$, so that
the rhs of Eq.~(\ref{logcauchybis}) must be multiplied by $M/q$.]
In agreement with the general discussion in Sec.~\ref{s:cumulants},
the position of the zero does not depend on the multiplicity $M$.
Equation~(\ref{flowestimate0}) then gives a spurious ``flow'' value
\begin{equation}
\label{spuriousinfty}
V_n^\theta\{\infty\}=q.
\end{equation}
It coincides with the limit of $V_n^\theta\{2k\}$ [Eq.~(\ref{spurious2k})]
for large $k$, as expected. The order of magnitude of this 
result is general if weights are of order unity.

Flow can be identified unambiguously only if it much
larger than the spurious value. 
Using Eq.~(\ref{Mvn}), and the assumption that $q$ is not 
much larger than unity, one finally obtains the condition 
under which our method can be applied: 
\begin{equation}
\label{vcondition}
v_n\gg \frac{1}{M}. 
\end{equation}
By comparison, the conditions under which one can apply 
the standard flow analysis (or two-particle methods) or 
four-particle cumulants are $v_n\gg 1/M^{1/2}$ and 
$v_n\gg 1/M^{3/4}$, respectively. The present method 
is more sensitive, in the sense that its validity extends 
down to smaller values of the flow. 

The condition (\ref{vcondition}) implies that the first zero of 
$G^\theta(z)$, approximately given by Eq.~(\ref{defz0}), 
satisfies the condition $|z_0|\ll 1$, which
justifies the approximation made at the beginning of
Sec.~\ref{s:Qdistribution}.
As already discussed in Refs.~\cite{Borghini:2000sa,Borghini:2001vi},
it is probably impossible to measure anisotropic flow
if this condition is not satisfied.

Note that the present model of nonflow correlations is a very extreme 
one, since we have assumed that they affect all particles,
and that particles within a jet are exactly collinear.
More realistic estimates of the spurious flow would give
significantly smaller values.
However, refining the above estimates would be academic:
we shall see in Sec.~\ref{s:spurious} that
if there is no anisotropic flow in the system,
the spurious flow created by statistical fluctuations
is much larger than the flow created by nonflow correlations,
unless the number of events is unrealistically large,
so that the present method will not be limited
by nonflow correlations in practice.

\section{Systematic errors}
\label{s:systematic}

In this Section, we estimate the order of magnitude of the systematic
error due to the method itself on our flow estimate.
We discuss first integrated flow, then differential flow. 
We assume that hypotheses C) and D)
are satisfied, i.e., we neglect statistical fluctuations
and detector effects.

\subsection{Integrated flow}
\label{s:systematic1}

The error on the integrated flow is a consequence of the two 
approximations made in Sec.~\ref{s:Qdistribution}.
The first one was to keep only the first two terms in the
power-series expansion of Eq.~(\ref{shortrange1}). The second one
was our assuming that $\sigma$ in Eq.~(\ref{defsigma2})
was independent of $\Phi_R$.

We first estimate the error resulting from the term proportional
to $z^3$ in Eq.~(\ref{shortrange1}), which was neglected in
Eq.~(\ref{centrallimit}).
We know from the general discussion in Sec.~\ref{s:hypotheses}
that the coefficient $c$ in front of $z^3$ is independent
of the size of the system, $M$, and vanishes when there
is no anisotropic flow. Hence, it is natural to assume
that it is of order $v_n$.
Since the value of $z$ of interest is of order $1/(M v_n)$
[from Eq.~(\ref{defz0})], the correction to the rhs of
Eq.~(\ref{centrallimit}) arising from the $z^3$ term is
of the order of $Mcz^3\sim 1/(Mv_n)^2$.
Since the lhs of Eq.~(\ref{centrallimit}) is
a logarithm, this is a {\em relative} correction to the
generating function. Hence, the {\em relative} error on
the flow estimate will be of the same order of magnitude.

Next, we estimate the error due to the dependence of $\sigma$ on $\Phi_R$.
A detailed calculation of this effect is performed in 
Appendix~\ref{s:systerrors}. 
In particular, it is shown that anisotropic flow results in contributions 
to $\sigma^2$ proportional to $\sin^2(n(\Phi_R-\theta))$.
These contributions consist of two terms, which are of magnitude $Mv_n^2$ 
and $Mv_{2n}$ (i.e., involving a higher harmonic), respectively.
Now, $\sigma^2$ is multiplied by $z^2$ in Eq.~(\ref{centrallimit}).
Replacing $z$ by the value of interest in Eq.~(\ref{defz0}),
$\Phi_R$-dependent terms yield contributions to the rhs of
Eq.~(\ref{centrallimit}) of orders $1/M$ and $v_{2n}/(Mv_n^2)$,
respectively.
As explained above, the relative error on the flow estimate due to 
these terms will be of the same order of magnitude.

Finally, note that the contribution of the term of order $z^4$
to the rhs of Eq.~(\ref{shortrange1}) may be of
the same order of magnitude, or even larger, than the
term of order $z^3$, although $|z|$ is much smaller
than unity. This is because the coefficient in front
of $z^3$ is of order $v_n$, which can be much smaller
than unity, while the coefficient in front of $z^4$
is of order unity. However, only the part of this
coefficient which depends on $\Phi_R$ matters:
indeed, the $\Phi_R$-independent part gives a multiplicative
contribution to $G^\theta(z)$ of the type $\exp(\lambda z^4)$,
which has no zeroes. Now, the $\Phi_R$-dependent contribution
to the term of order $z^4$ is much smaller than the
$\Phi_R$-contribution to the term of order $z^2$,
which has been estimated above.

In summary, the order of magnitude of the relative systematic
error on the flow is
\begin{eqnarray}
\label{systerror}
\frac{V_n\{\infty\}-V_n}{V_n} &=&{\cal O}\!\left(\frac{1}{M}\right)+
{\cal O}\!\left(\frac{1}{(M v_n)^2}\right)\cr
& & +\ {\cal O}\!\left(\frac{v_{2n}}{M v_n^2}\right).
\end{eqnarray}
All three terms involve powers of $1/M$: they can be viewed
as finite multiplicity corrections, resulting from the violation
of hypothesis A) in Sec.~\ref{s:hypotheses}.

Let us comment on the magnitude of these errors.
The first term, a relative correction of order $1/M$,
is always small since $M\gg 1$.
The last two terms are much less trivial.
They also appear when flow is analyzed from the cumulant
of four-particle correlations, where they result from
an interference between correlations due to flow and
nonflow correlations, as discussed in detail in Appendix A in
Ref.~\cite{Borghini:2000sa}.
As long as condition (\ref{vcondition}) is satisfied, the term
of order $1/(Mv_n)^2$ is a small correction.
The relative magnitude of the first two error terms in Eq.~(\ref{systerror})
actually depends on the value of the resolution parameter $\chi$
which will be defined below in Sec.~\ref{s:resolution},
and is of order $v_n\sqrt{M}$.
The largest correction term in Eq.~(\ref{systerror}) is the first one for
large $\chi$, and the second one for small $\chi$.
Higher harmonics are generally of smaller magnitude: typically,
$v_{2n}\sim v_n^2$, so that the third term in the rhs of
Eq.~(\ref{systerror}) is comparable to the first term.
An important exception is that directed flow is much
smaller than elliptic flow at ultrarelativistic energies.
In this regime, the third term dominates the error for $n=1$.

Nonflow correlations involving a few particles increase the 
systematic error on the flow, but the order of magnitude
is still given by Eq.~(\ref{systerror}). In the extreme
case where all particles are emitted in bunches of $q$ collinear
particles, as in Sec.~\ref{s:sensitivity},
one must replace in Eq.~(\ref{systerror}) $M$ by $M/q$, which
is the effective multiplicity, i.e., the number of independent angles
(the number of independent degrees of freedom).
Therefore the first and third
term in Eq.~(\ref{systerror}) will be increased by a factor
of $q$, and the second term by a factor of $q^2$.

Finally, how does our systematic error compare with the systematic
error on the flow estimate $V_n\{2k\}$ from cumulants of finite
order $2k$?
Possible nonflow correlations between $2k$ particles yield
an additional term:
\footnote{The magnitude of the systematic error was given correctly in 
Appendix~A of Ref.~\cite{Borghini:2000sa}, but not in 
Ref.~\cite{Borghini:2001vi} where only the last term was kept, which was a 
mistake.}
\begin{eqnarray}
\label{systerror2k}
\frac{V_n\{2k\}-V_n}{V_n} &=&{\cal O}\!\left(\frac{1}{M}\right)+
{\cal O}\!\left(\frac{1}{(M v_n)^2}\right)\cr
& & +\ {\cal O}\!\left(\frac{v_{2n}}{M v_n^2}\right)
+{\cal O}\!\left(\frac{1}{M^{2k-1} v_n^{2k}}\right)\!.\qquad
\end{eqnarray}
%The last term becomes negligible as soon as
%\begin{equation}
%\label{drosophile}
%v_n>\frac{1}{M^{1-1/(2k-2)}},
%\end{equation}
%but dominates if $v_n$ is smaller than this value.
In particular, this equation can be used to show that the error on flow 
estimates from standard (event-plane or two-particle) methods is much 
larger than with the present method. 
Setting $2k=2$ in the previous equation, the last term dominates over the
first three terms, so that we may write
\begin{equation}
\frac{V_n\{2\}-V_n}{V_n}=\mathcal{O}\left(\frac{1}{Mv_n^2}\right).
\end{equation}
With higher-order cumulants, $2k\ge 4$, the last term in 
Eq.~(\ref{systerror2k}) dominates only if
\begin{equation}
\label{drosophile}
v_n<\frac{1}{M^{1-1/(2k-2)}}.
\end{equation}
If $v_n$ is larger than this value, the error is similar with the cumulant 
method and with the new method presented here.

\subsection{Differential flow}
\label{s:systematic2}

The determination of differential flow relies on the previous 
determination of integrated flow: one thus naturally 
expects a relative error on the differential flow of the 
same order as the relative error on the integrated flow, 
Eq.~(\ref{systerror}). This can be checked explicitly, see 
Appendix~\ref{s:systerrors}.

An additional systematic error is due to the approximation made
in writing Eq.~(\ref{centrallimit2}). It was assumed that the
proton and the flow vector are independent. However, they may 
be correlated, due to autocorrelations (if the proton is 
involved in the flow vector) or to nonflow correlations, whose
effects are always similar to those of  autocorrelations 
(see Sec.~\ref{s:autocorrelations}). 

We now compute the error due to autocorrelations. For
this purpose, we compute the next-to-leading term $b'z$ in 
the power-series expansion Eq.~(\ref{shortrange2}).
We compute only the contribution of the numerator of the 
lhs to this term. The contribution of the denominator yields a 
relative error on $v'_{mn}$ of the same order as for the
integrated flow, Eq.~(\ref{systerror}). 
If the proton has weight $w'$ in the flow vector, a simple 
calculation gives
\begin{eqnarray}
b'&=&\frac{w'}{2}\left(v'_{(m-1)n}\cos[(m-1)n(\Phi_R-\theta)]\right.\cr
& &\qquad \left. +\,v'_{(m+1)n}\cos[(m+1)n(\Phi_R-\theta)]\right).
\end{eqnarray}
Taking this term into account in Eq.~(\ref{centrallimit2}) and 
the following equations, $D^\theta_m(z)$ is no longer 
given by Eq.~(\ref{dmcl}) but by 
\begin{widetext}
\begin{equation}
\label{dmbcl}
D^\theta_{m}(z) = e^{\sigma^2 z^2/4} \left[I_m(V_n z)\,v'_{mn} + 
\frac{z}{2}\,I_{m-1}(V_n z)\,w'v'_{(m-1)n} +
\frac{z}{2}\,I_{m+1}(V_n z)\,w'v'_{(m+1)n}\right].
\end{equation}
\end{widetext} 
Since higher harmonics are generally of smaller magnitude, 
we neglect the last term in the rhs. 
Evaluating the remaining two terms at $z=ir_0=ij_{01}/V_n$, the differential 
flow estimate $v^{\prime}_{mn}\{\infty\}$ is given by 
\begin{equation}
\label{systauto}
v^{\prime}_{mn}\{\infty\} = v^{\prime}_{mn} +
\frac{j_{01}J_{m-1}(j_{01})}{2J_m(j_{01})}
\frac{w'v'_{(m-1)n}}{V_n}.
\end{equation}
For $m=1$, the extra correction term vanishes since $J_0(j_{01})=0$. 
One can show that the next term $c'z^2$ in the expansion 
Eq.~(\ref{shortrange2}) produces a relative error on $v'_n$
of the same order as that on the integrated flow,
Eq.~(\ref{systerror}). 
For $m=2$, on the other hand, we obtain 
\begin{equation}
\label{systhigher}
v^{\prime}_{2n}\{\infty\}=v^{\prime}_{2n}
+\frac{j_{01}^2}{4}\frac{w'v'_{n}}{V_n}.
\end{equation}
Numerically with the input values $M=300$ particles, $v_2=6\%$, 
$v'_2=7\%$, $v'_4=3\%$ 
used for the simulation of Secs.~\ref{s:recipeint} and \ref{s:recipedif}, 
we find $v^{\prime}_4\{\infty\}=3.56\%$, in agreement with 
the result of the analysis when autocorrelations are not removed, 
$v'_4\{\infty\}=3.60\pm 0.29\%$,  see Fig.~\ref{fig:vdiff}. 

As a conclusion, the systematic error on the differential flow 
in the lowest harmonic $v'_n$ is a relative error of the same 
order of magnitude as for the integrated flow: it is small, and 
autocorrelations need not be removed. 
The situation is quite different for the higher harmonics
$v'_{2n}$, $v'_{3n}$..., where autocorrelations give a sizeable
contribution. This contribution can be removed analytically 
using Eq.~(\ref{systhigher}). 
However, one must be aware that an error of the same order 
of magnitude may remain as a result of nonflow correlations.
The same conclusion holds with other methods of flow analysis. 

This error on higher harmonics has two consequences. 
At ultrarelativistic energies, where the flow vector is constructed
with $n=2$, the higher harmonic $v_4$ is expected to be 
of much smaller magnitude~\cite{Kolb:1999it}. Since, furthermore,
nonflow correlations are known to be 
sizeable~\cite{NA49new,Adler:2002pu}, one may doubt whether 
$v_4$ can be measured at all. 
At lower energies, elliptic flow $v_2$ is usually measured using the 
flow vector from directed flow 
$n=1$~\cite{Demoulins:ac,Andronic:2000cx,Chung:2001qr}, this bias
should be considered seriously. In particular, it may be large 
for mesons, which often originate from resonance decays and have
therefore strong nonflow correlations.

\section{Flow fluctuations}
\label{s:fluctuations}

As explained in Sec.~\ref{s:hypotheses},
our results rely on the hypothesis that we are considering
events with exactly the same impact parameter.
In practice, however, the set of events in a given
centrality bin spans some impact parameter interval.
In this Section, we are going to study the influence of
such impact parameter fluctuations on the flow
analysis~\cite{Adler:2002pu}.

We assume that the hypotheses of Sec.~\ref{s:hypotheses} are
satisfied for a fixed impact parameter $b$, and we denote
the integrated flow at this impact parameter by $V_n(b)$.
If $b$ fluctuates in the sample of events, one can compute
the generating function $G^\theta(ir)$ in two steps.
One first averages over events with the same impact parameter $b$:
$\smean{e^{irQ^\theta}|b}$ is given by the rhs of
Eq.~(\ref{meanGflow}), where both $V_n$ and $\sigma$ may
depend on $b$. Then, one averages over $b$:
\begin{equation}
\label{fluct0}
G^\theta(ir)=\mean{J_0(V_n(b) r)\,e^{-\sigma(b)^2 r^2/4}}_b, 
\end{equation}
where angular brackets denote an average over $b$.

Our flow estimate $V_n\{\infty\}$ is related to the first root,
$r_0^\theta$, of the equation $G^\theta(ir_0^\theta)=0$,
by Eq.~(\ref{flowestimate0}).
First of all, if only $\sigma(b)$ fluctuates, while $V_n(b)$ is
a constant, the position of the zero does not depend on $b$: this
shows that the estimate of the flow is not affected by
fluctuations of $\sigma$ if only those are present.
Next, if $V_n(b)$ has small fluctuations around an average
value, i.e., $V_n(b)=\bar V_n+\delta V_n(b)$, then expanding
Eq.~(\ref{fluct0}) to first order in $\delta V_n(b)$,
one easily shows that $V_n\{\infty\}=\bar V_n$, that is,
the analysis using the present method yields the {\em average value} 
of the flow.

When fluctuations are large, our estimate $V_n\{\infty\}$ does not
in general coincide with the average value of the flow, $\bar V_n$.
It is sensitive to the fluctuations of $V_n(b)$, and also to the
fluctuations of the width $\sigma(b)$, which is involved in
Eq.~(\ref{fluct0}).
However, it is possible to show under rather general conditions
that if $V_n(b)$ lies between $(V_n)_{\rm min}$ and $(V_n)_{\rm max}$,
then $V_n\{\infty\}$ also lies between
$(V_n)_{\rm min}$ and $(V_n)_{\rm max}$.
For this purpose, we have to show that the first zero
of $G^\theta(ir)$ defined by Eq.~(\ref{fluct0}) occurs
between $r=j_{01}/(V_n)_{\rm max}$ and $r=j_{01}/(V_n)_{\rm min}$.
Since $J_0(x)>0$ for $0<x<j_{01}$,
and $V_n(b)<(V_n)_{\rm max}$,
Eq.~(\ref{fluct0}) shows that $G^\theta(ir)>0$
for $r<j_{01}/(V_n)_{\rm max}$.
To prove our result, it is sufficient to show that
$G^\theta(ij_{01}/(V_n)_{\rm min})<0$.
Using Eq.~(\ref{fluct0}), this holds as soon as
$J_0(j_{01}V_n(b)/(V_n)_{\rm min})<0$ for all $b$.
Now, $J_0(x)$ is negative for $x$ between
$j_{01}$ and its second zero, $j_{02}$.
Hence our result holds as soon as
\begin{equation}
j_{01}<j_{01}\,\frac{V_n(b)}{(V_n)_{\rm min}}<j_{02}
\end{equation}
for all $b$. One obtains the condition:
\begin{equation}
\frac{(V_n)_{\rm max}}{(V_n)_{\rm min}} < \frac{j_{02}}{j_{01}} \simeq 2.3.
\end{equation}
This condition is sufficient but not necessary.
It shows that if the flow does not vary by a factor of more
than 2.3, the flow estimate $V_n\{\infty\}$ lies between 
$(V_n)_{\rm min}$ and $(V_n)_{\rm max}$, which is quite reasonable.

\section{Statistical errors}
\label{s:statistical}

In this Section, we estimate the statistical errors which arise when
the number of events $\N$ is finite.
As with other methods of flow analysis, these statistical
errors depend on the resolution parameter $\chi$.
In Sec.~\ref{s:resolution}, we define this parameter, explain
how it can be measured experimentally, and give its approximate
value for a variety of heavy-ion experiments.

We then study, in Sec.~\ref{s:fluctsample},
the fluctuation of the generating function $G^\theta(ir)$, evaluated
on the imaginary axis (i.e., for real $r$), around the true
statistical average
$G_{\rm c.l.}(ir)$. Then, in Sec.~\ref{s:spurious}, we
show that even if there is no flow in the system,
the data analysis through the present method will yield a non-zero 
``flow'' value, which
we estimate: we consider it as the lower bound on
the flow, below which the present method cannot be used.
When there is flow in the system, and when it is larger
than this limiting value, statistical errors are small.
They are estimated in Sec.~\ref{s:statint} for the
integrated flow and in Sec.~\ref{s:statdiff} for the
differential flow.

\subsection{The resolution parameter}
\label{s:resolution}

An important parameter in the flow analysis is the
resolution parameter $\chi$ defined as
\footnote{Please note that this definition of $\chi$ differs by a factor of
$1/\sqrt{2}$ from the one adopted in Ref.~\cite{Poskanzer:1998yz}.}
\begin{equation}
\label{defchi}
\chi\equiv \frac{V_n}{\sigma},
\end{equation}
where $V_n$ and $\sigma$ are defined in Eqs.~(\ref{defVn}) and
(\ref{defsigma2}), respectively.
Physically, it characterizes the relative strength of the flow
compared to the finite-multiplicity fluctuations.

With unit weights, $V_n=Mv_n$ [Eq.~(\ref{Mvn})] and
$\sigma=\sqrt{M}$ for independent particles [Eq.~(\ref{sigma2})],
so that $\chi=v_n\sqrt{M}$.
This shows that $\chi$ increases with both the flow and the number of
detected particles.
It is small for central collisions where $v_n$ is small due to azimuthal
symmetry, and also for peripheral collisions where the multiplicity
$M$ is small.
It is generally maximum for intermediate centralities.
Here again, we want to emphasize that one should use all detected particles 
in the analysis, so as to maximize $\chi$. 

This parameter is related to the so-called ``event-plane resolution''
in the standard flow analysis.
The event-plane resolution is defined as the average value of
$\cos\Delta\Phi_R\equiv \cos(n(\Phi_n-\Phi_R))$, where $n\Phi_n$ is the
azimuthal angle of the flow vector and $\Phi_R$ that of the reaction plane.
It is related to $\chi$ by~\cite{Ollitrault:1997di}:
\begin{equation}
\label{pn}
\langle\cos\Delta\Phi_R\rangle = \frac{\sqrt{\pi}}{2}\chi e^{-\chi^2/2}
\left[ I_0\!\left( \frac{\chi^2}{2}\right) +
I_1\!\left( \frac{\chi^2}{2}\right) \right].
\end{equation}

In Table~\ref{Tchi}, we have summarized the values of $\chi$ for various 
heavy-ion experiments in mid-central collisions (where $\chi$ is largest),
calculated with the help of Eq.~(\ref{pn}) using the values of the
event-plane resolution given in the cited references.
These values of $\chi$ lie typically between 0.5 and 2.5.
Note, however, that these values may not reflect only flow, but could be 
contaminated by nonflow effects. 
Such an instance, namely the influence of nonflow correlations from global 
momentum conservation on the determination of the first-harmonic event-plane 
resolution, was discussed in Ref.~\cite{Borghini:2002mv}. 

\begin{table}[ht]
\caption{Values of the resolution parameter $\chi$ and the ``dispersion'' 
factor used in the standard flow analysis for various experiments. 
$n$ is the Fourier harmonic used to determine the event plane (1 for directed 
flow, 2 for elliptic flow).}
\label{Tchi}
\begin{center}
\begin{tabular}{|c|c|c|c|c|c|c|}
\hline
$\chi$ &$\mean{\cos(\Delta\Phi_R)}$& $n$& System & $E/A$ & Exp.
& Ref.\\
\hline
0.90 & 1/1.56 & $1$ & Au+Au & 90~MeV &FOPI & \cite{Andronic:2001sw}\\
\hline
2.6 & 1/1.04 & $1$ &Au+Au & 400~MeV &FOPI& \cite{Andronic:2001sw}\\
\hline
1.65 & 0.89 & $1$ &Au+Au &2~GeV &E895& \cite{Liu:2000am}\\
\hline
0.52 & 0.43 & $1$ &Au+Au & 8~GeV &E895& \cite{Liu:2000am}\\
\hline
0.39 & 0.33 & $1$ &Pb+Pb & 158~GeV &NA49& \cite{NA49new}\\
\hline
0.67 & 0.53 & $2$ &Pb+Pb & 158~GeV & NA49&\cite{NA49new}\\
\hline
1.25 & 0.8 & $2$ &Au+Au & 65+65~GeV &STAR& \cite{Adler:2002pu}\\
\hline
0.48 & 0.4 & $2$ &Au+Au & 100+100~GeV &PHENIX& \cite{PHENIX200}\\
\hline 
\end{tabular}
\end{center}
\end{table}

How can $\chi$ be determined experimentally?
The numerator of Eq.~(\ref{defchi}) is the integrated flow $V_n$, which 
is determined following the procedure introduced in Sec.~\ref{s:recipeint}.
The denominator, $\sigma$, is most simply obtained by averaging
Eq.~(\ref{defsigma2}) over $\theta$.
In order to take into account possible azimuthal asymmetries in
the detector acceptance, which were neglected in Sec.~\ref{s:Qdistribution},
we replace Eq.~(\ref{averageQ}) with
\begin{eqnarray}
\label{averageQbis}
\mean{Q^\theta|\Phi_R}&=&V_n\,\cos(n(\Phi_R-\theta))\cr
& & +\ \mean{Q_x}\cos(n\theta)+\mean{Q_y}\sin(n\theta),
\end{eqnarray}
where we have used the first line of Eq.~(\ref{defqtheta}).
The two additional terms vanish with a symmetric detector.
One then obtains
\begin{equation}
\label{sigmaexp2}
\sigma^2=\mean{Q_x^2+Q_y^2}-\mean{Q_x}^2-\mean{Q_y}^2-V_n^2.
\end{equation}
The value of $\sigma$ obtained in this way is more accurate
than the approximate value in Eq.~(\ref{sigma2}), which
underestimates $\sigma$ when large nonflow
correlations are present.
Since statistical errors are very sensitive to $\chi$, as we
shall see, it is important to estimate $\sigma$ as accurately
as possible, in a model independent way, which is possible
using Eq.~(\ref{sigmaexp2}).

\subsection{Preliminaries}

Here we derive simple results which will be useful in the
remainder of the Section.
The estimate of the integrated flow, $V_n\{\infty\}$, is obtained by 
averaging $V_n^\theta\{\infty\}$ over several values of $\theta$, see 
Eq.~(\ref{av_over_theta}), and a similar equation for differential flow:
\begin{equation}
v'_{mn}\{\infty\}\equiv
\frac{1}{p}\sum_{k=0}^{p-1}v_{mn}^{\prime k\pi/np}\{\infty\}.
\end{equation}
In order to estimate the standard deviation of the average
$V_n\{\infty\}$, one needs to evaluate not only the standard deviation of
$V_n^\theta\{\infty\}$, but also the correlation between
different estimates $V_n^\theta\{\infty\}$ and 
$V_n^{\theta'}\{\infty\}$ with $\theta\neq\theta'$.
For this purpose, we shall have to evaluate averages such as 
$\smean{e^{irQ^\theta} e^{ir'Q^{\theta'}}}$.
We introduce the complex notation
\begin{equation}
\label{complex0}
z=r e^{in\theta}, \qquad z'=r' e^{in\theta'}.
\end{equation}
One should note that the number $z$ thus defined has nothing to do with
the variable $z$ involved in Eq.~(\ref{defGint}).
With this notation, we have
\begin{equation}
\label{complex1}
e^{ir Q^\theta} = e^{i(z^*Q+zQ^*)/2},
\end{equation}
where $Q\equiv Q_x+iQ_y$. Then, obviously,
\begin{equation}
\label{complex2}
e^{ir Q^\theta}e^{ir' Q^{\theta'}}=e^{i[(z^*+z^{\prime *})Q+(z+z')Q^*]/2}.
\end{equation}
We assume that the expectation value of $e^{ir Q^\theta}$
is given by our theoretical estimate, Eq.~(\ref{meanGflow}).
Then,
\begin{eqnarray}
\label{g2}
\mean{e^{ir Q^\theta}e^{ir' Q^{\theta'}}}&=&G_{\rm c.l.}(i|z+z'|)\cr
\mean{e^{ir Q^\theta}e^{-ir' Q^{\theta'}}}&=&G_{\rm c.l.}(i|z-z'|),
\end{eqnarray}
from which we obtain
\begin{widetext}
\begin{eqnarray}
\label{ReImG}
\mean{{\rm Re}\,(e^{irQ^\theta})\,
{\rm Re}\,(e^{ir'Q^{\theta'}})}&=&
\frac{G_{\rm c.l.}(i|z-z'|)+G_{\rm c.l.}(i|z+z'|)}{2}\cr
\mean{{\rm Im}\,(e^{irQ^\theta})\,
{\rm Im}\,(e^{ir'Q^{\theta'}})}&=&
\frac{G_{\rm c.l.}(i|z-z'|)-G_{\rm c.l.}(i|z+z'|)}{2}.
\end{eqnarray}
\end{widetext}
All these identities will be used below.

\subsection{Fluctuations of $G^\theta(ir)$}
\label{s:fluctsample}

In this Section, we estimate the magnitude of the statistical fluctuations 
of (the complex-valued) function $G^\theta(ir)$ around its (real-valued) 
average $G_{\rm c.l.}(ir)$. 
In particular, we derive the fluctuations of the modulus $|G^\theta(ir)|$, 
which is the quantity minimized in Sec.~\ref{s:recipeint} to extract integrated 
flow. 

Let us first recall general results about sample averages,
i.e., average values evaluated from a finite sample of events.
If $x$ is an observable measured in an
event (multiplicity, transverse energy, etc.), we denote by $\sample{x}$
the average value of $x$ over the available sample of events, which is
also called the {\it sample average\/}:
\begin{equation}
\label{sampling}
\sample{x}\equiv\frac{1}{\N}\sum_{\alpha=1}^{\N} x_\alpha.
\end{equation}
The {\em expectation value}, which is the limit of this quantity
when the number of events goes to infinity,
is denoted by angular brackets as in the previous Sections.
Note that $\mean{\sample{x}} = \mean{x}$.

If $\N$ is large but still finite, the sample average differs from
the expectation value by a small fluctuation $\delta x$:
$\sample{x}=\mean{x}+\delta x$.
If $y$ denotes another observable associated with each event, then
the covariance of the sample averages $\sample{x}$ and
$\sample{y}$ is
\begin{eqnarray}
\label{samplingcov}
\mean{\delta x\,\delta y}&\equiv&
\mean{\sample{x}\sample{y}}-\mean{x}\mean{y}\cr
&=&\frac{1}{\N}\left(\mean{xy}-\mean{x}\mean{y}\right),
\end{eqnarray}
where in the second line we have used the property that events are 
statistically independent.
Note that the statistical properties of the fluctuations
$\delta x$, $\delta y$
are Gaussian due to the central limit theorem, hence they are
completely determined by their covariance: higher-order
moments are obtained using Wick's theorem.

We can apply Eq.~(\ref{samplingcov}) to compute the statistical fluctuations
of the generating function, Eq.~(\ref{defGint}), for $z$ on the imaginary axis 
($z=ir$, with $r$ real). 
We introduce the decomposition
\begin{equation}
G^\theta(ir)=G_{\rm c.l.}(ir)+\delta G^\theta(ir),
\end{equation}
where $\delta G^\theta(ir)$ denotes the fluctuation of $G^\theta(ir)$ around 
its expectation value.

Let us first estimate the magnitude of $\delta G^\theta(ir)$.
For this purpose, we make use of Eq.~(\ref{samplingcov}) with 
$x=e^{irQ^\theta}$ and $y=e^{-irQ^\theta}$:
\begin{equation}
\label{Absdelta}
\mean{|\delta G^\theta(ir)|^2}
=\frac{1}{\N}\left(1-G_{\rm c.l.}(ir)^2\right).
\end{equation}
At the zero of $G_{\rm c.l.}(ir)$, the standard deviation of 
$|G^\theta(ir)|$ is exactly $1/\sqrt{\N}$. 
This justifies Eq.~(\ref{zerocheck}). 
For large $r$, the expectation value $G_{\rm c.l.}(ir)$ goes to zero 
[see Eq.~(\ref{meanGflow})], while the fluctuations do not:
$|\delta G^\theta(ir)|$ is of order $1/\sqrt{\N}$.
This can be easily understood: for large $r$, 
$e^{ir Q^\theta}$ is a random complex number
uniformly distributed on the unit circle, so that
the sample average $G^\theta(ir)$ in Eq.~(\ref{defqtheta})
is a random walk of $\N$ steps, of length $1/\N$ each, hence the 
order of magnitude of the fluctuations. 

We can now estimate the value of $r$ for which the fluctuation
$\delta G^\theta(ir)$ becomes of the same order of magnitude as the
expectation value $G_{\rm c.l.}(ir)$. The decrease of $G_{\rm c.l.}(ir)$
for large $r$ in Eq.~(\ref{meanGflow}) is dominated by
the exponential factor, while Eq.~(\ref{Absdelta}) shows that 
$|\delta G^\theta(ir)|$ is of order $1/\sqrt{\N}$, so that both become 
comparable when
\begin{equation}
\frac{1}{\sqrt{\N}}\simeq e^{-\sigma^2 r^2/4}.
\end{equation}
We denote by $r_c$ the corresponding value of $r$:
\begin{equation}
\label{zcrit}
r_c\equiv\frac{1}{\sigma}\sqrt{2\ln\N}.
\end{equation}
Fluctuations are relatively small below the critical value $r_c$ and
dominate above. In the case of the Monte-Carlo simulation presented
in Fig.~\ref{fig:Gvsr}, for instance, taking $\sigma=\sqrt{M}$, one
obtains $r_c\simeq 0.26$: it is larger by a factor of $2$ than the
position of the first minimum, $r_0^\theta$, which is the reason
why the statistical fluctuations on our flow estimates, shown
in Fig.~\ref{fig:v2theta}, are small.

We finally estimate the fluctuations of the modulus
$\left|G^\theta(ir)\right|$, which is involved
in determining the integrated flow (see Sec.~\ref{s:recipeint}).
We assume that $r<r_c$, so that $\left|\delta G^\theta(ir)\right|$ is 
much smaller than $G_{\rm c.l.}(ir)$.
We separate the fluctuation into its real and imaginary part:
\begin{equation}
\label{ReImdeltaG}
G^\theta(ir)=G_{\rm c.l.}(ir)+
{\rm Re}\,\delta G^\theta(ir)+i\,{\rm Im}\,\delta G^\theta(ir).
\end{equation}
Taking the modulus of Eq.~(\ref{ReImdeltaG}),
we then obtain, to leading order in $\delta G^\theta(ir)$
\begin{equation}
\label{Imneglected}
|G^\theta(ir)|\simeq\left|G_{\rm c.l.}(ir)+{\rm Re}\,\delta G^\theta(ir)\right|,
\end{equation}
i.e., only the real part of the fluctuations of $G^\theta(ir)$ contribute to 
the fluctuations of the modulus $|G^\theta(ir)|$.
Their covariance is obtained from Eq.~(\ref{samplingcov}), in which we take
$x={\rm Re}\,\delta G^\theta(ir)$ and
$y={\rm Re}\,\delta G^{\theta'}(ir')$, and with the help of Eq.~(\ref{ReImG}):
\begin{widetext}
\begin{equation}
\label{RedeltaG}
\mean{{\rm Re}\,\delta G^\theta(ir)\,{\rm Re}\,\delta G^{\theta'}(ir')}
=\frac{1}{2\N}\left[G_{\rm c.l.}(i|z-z'|)+G_{\rm c.l.}(i|z+z'|)-2\,
G_{\rm c.l.}(ir)\,G_{\rm c.l.}(ir')\right],
\end{equation}
\end{widetext}
where we have used the property that the expectation value
$G_{\rm c.l.}(ir)$ is real [see Eq.~(\ref{meanGflow})].

\subsection{Spurious flow from statistical fluctuations}
\label{s:spurious}

Let us now assume that there is no anisotropic flow in the system,
$V_n=0$. 
In that case, analyzing the data with the present method will nevertheless 
yield a spurious ``flow'' value, due to statistical fluctuations. 
In this Section, we first estimate the maximal such spurious flow value 
which analyses would give in the absence of real flow. 
This maximal value is in our eyes the minimal flow value which the method can 
safely reconstruct. 
Then we show that any $V_n$ extracted with the method larger than this 
value can in turn be reliably attributed to flow. 

If there is no flow in the system, Eq.~(\ref{meanGflow}) reads
\begin{equation}
\label{meanGnoflow}
G_{\rm c.l.}(i r)=e^{-\sigma^2 r^2/4},
\end{equation}
so that there is no zero, nor any minimum of the true expectation
value $G_{\rm c.l.}(i r)$ for finite $r$.
With a finite number of events, however, $\left|G^\theta(i r)\right|$ 
does in general have a minimum, so that the procedure outlined in 
Sec.~\ref{s:recipeint} will yield a non-zero estimate of the flow, 
$V_n^\theta\{\infty\}$, which is unphysical.

This minimum is due to the fluctuation $\delta G^\theta(ir)$:
it typically occurs when $\delta G^\theta(ir)$ is of the same order
of magnitude as the average value $G_{\rm c.l.}(ir)$.
The position of the minimum, $r_0^\theta$, is therefore of
order $r_0^\theta\sim r_c$, where $r_c$ is given by Eq.~(\ref{zcrit}).
The resulting ``spurious flow'' is given by Eq.~(\ref{flowestimate0}).
To this spurious flow, we can associate a resolution parameter as in 
Eq.~(\ref{defchi}):
\begin{equation}
\label{chitheta}
\chi^\theta\equiv \frac{V_n^\theta\{\infty\}}{\sigma}.
\end{equation}
[Experimentally, $\sigma$ should be evaluated using 
Eq.~(\ref{sigmaexp2}) with $V_n=0$.] 
Using Eqs.~(\ref{flowestimate0}) and (\ref{zcrit}), this yields
\begin{equation}
\label{spurious}
\chi^\theta\sim\frac{j_{01}}{\sqrt{2\,\ln\,\N}}.
\end{equation}
With $\N=20000$~events, one obtains numerically $\chi^\theta=0.54$.
If the analysis yields this (or a smaller) value for $\chi$, the result 
cannot be attributed to flow, and is just an artifact due to statistical 
fluctuations. 
Experiments for which the resolution parameter $\chi$ (see the values 
in Table~\ref{Tchi}) is not larger than this spurious value will not 
have enough statistics to implement the present method, unless a proper 
choice of weights can significantly increase the value of $\chi$. 

The values of $\chi$ for actual heavy-ion experiments,
listed in Table~\ref{Tchi}, are often larger than this value,
but not much larger, which might look somewhat worrying at
first sight. Fortunately, as soon as the analysis yields a value
slightly higher than $\chi^\theta$ in Eq.~(\ref{spurious}),
the result can safely be attributed to collective flow, as we shall now 
show. 

As explained in Sec.~\ref{s:fluctsample}, a minimum of $|G^\theta(ir)|$ 
occurs approximately when the real part of $G^\theta(ir)$ vanishes, i.e., 
when $-{\rm Re}\,\delta G^\theta(ir)=G_{\rm c.l.}(ir)$, see 
Eq.~(\ref{Imneglected}). 
Now, ${\rm Re}\,\delta G^\theta(ir)$ has Gaussian fluctuations,
whose width is given by Eq.~(\ref{RedeltaG}), in which we set $z=z'$, and 
neglect the last two terms in the brackets in the rhs (which are at most of 
order $1/\sqrt{\N}$), so that
\begin{equation}
\mean{\left({\rm Re}\,\delta G^\theta(ir)\right)^2}
=\frac{1}{2\N}.
\end{equation}
We can say with 98\% confidence level that
$-{\rm Re}\,\delta G^\theta(ir)$ is smaller than
twice this standard deviation, i.e.,
\begin{equation}
-{\rm Re}\,\delta G^\theta(ir) <\sqrt{\frac{2}{\N}}.
\end{equation}
Therefore, if a minimum of $|G^\theta(ir)|$ occurs at
$r=r_0^\theta$, we can say with the same confidence level that
\begin{equation}
G_{\rm c.l.}(i r_0^\theta) <\sqrt{\frac{2}{\N}}.
\end{equation}
Using Eqs.~(\ref{flowestimate0}), (\ref{meanGnoflow}) and
(\ref{chitheta}), we obtain the following inequality
with 98\% confidence level:
\begin{equation}
\label{upperbound}
\chi^\theta<\frac{j_{01}}{\sqrt{2\,\ln\,(\N/2)}}.
\end{equation}
With the same value $\N=20000$ as above, the upper bound
is $0.56$, only very slightly larger than the estimate
from Eq.~(\ref{spurious}).
This shows that even a value of the resolution parameter
slightly larger than the rhs of Eq.~(\ref{spurious})
cannot be attributed to statistical fluctuations alone.
In addition, this discussion holds for a fixed value of
the reference angle $\theta$.
Averaging over several values of $\theta$ decreases
the statistical error, and also slightly decreases
the value of the spurious flow.
This is why we gave $\chi=0.5$ as the limiting value below which the 
method cannot be applied reliably in Sec.~\ref{s:recipestat}. 

\begin{center}
\begin{figure}[ht!]
\centerline{\includegraphics*[width=\linewidth]{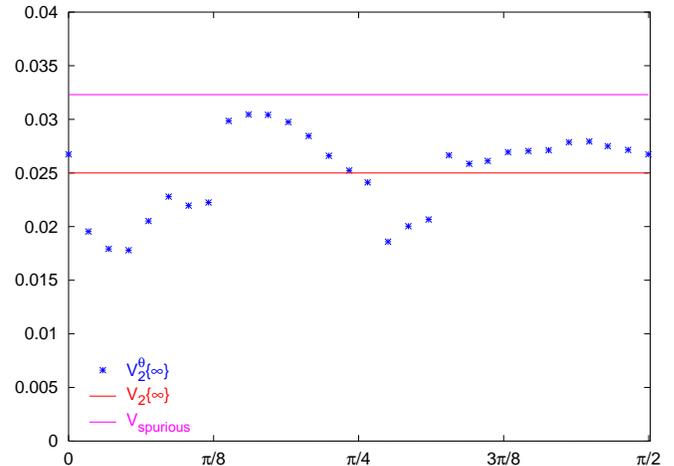}}
\caption{Analysis of $\N=20000$ events of multiplicity $M=300$,
simulated without anisotropic flow. Crosses show the value of the 
spurious flow given by our analysis, $V_2^\theta\{\infty\}/M$, as a 
function of $\theta$. The solid line displays the 98\% CL upper bound 
on this spurious flow (see text).}
\label{fig:Vspurious}
\end{figure}
\end{center}

In order to illustrate this discussion, we have simulated a
data set with $\N=20000$ events of multiplicity $M=300$ each.
Events were simulated with zero elliptic flow, $v_2=0$ for all particles, 
and we assumed a perfectly isotropic detector.
The data set was then treated according to the procedure
presented in Sec.~\ref{s:recipes}, using unit weights
$w_j=1$ and $n=2$ in Eq.~(\ref{flowvector}).
The resulting $V_2^\theta\{\infty\}/M$ is displayed in
Fig.~\ref{fig:Vspurious} as a function of $\theta$,
together with the upper bound given by Eqs.~(\ref{chitheta})
and (\ref{upperbound}), in which we set $\sigma=\sqrt{M}$.
One first notes that all estimates $V_2^\theta\{\infty\}$ are smaller 
than the bound, which means at once that they should not be attributed 
to flow. The average over $\theta$, 
$V_2\{\infty\}/M=2.56\%$, is well below the upper bound.
In addition, the values of $V_2^\theta\{\infty\}$ strongly depend on $\theta$, 
while we recall that in Fig.~\ref{fig:v2theta}, also obtained in the case of 
a perfect detector, they were roughly independent of $\theta$. 

Finally, a close look at the figure shows that the variation of
$V_2^\theta\{\infty\}$ with $\theta$ has four discontinuities, which can 
be qualitatively understood in the following way: when $\theta$ varies, 
$G^\theta(ir)$ varies continuously, but the first minimum of $|G^\theta(ir)|$ 
may disappear for some value of $\theta$.
Then, the second minimum becomes the first, and $V_2^\theta\{\infty\}$
suddenly jumps to a lower value (and vice-versa when
a minimum appears below the first minimum).
Such discontinuities are not expected when the minimum
is due to anisotropic flow, and are by themselves a hint
that the observed flow estimate is an effect of statistical 
fluctuations.
\footnote{The appearance of such discontinuities caused by statistical 
fluctuations, which do not occur for collisions with real flow, may be a 
good way to discriminate between both effects in the case of weak flow, 
slightly above the limit Eq.~(\ref{spurious}), provided one computes 
estimates $V_n^\theta\{\infty\}$ for a large number of values of $\theta$, 
instead of the 4 or 5 values which are enough to guarantee ``low'' 
statistical errors as argued in Sec.~\ref{s:statint}.}

\subsection{Statistical error on the integrated flow}
\label{s:statint}

In this Section, we again assume that there is flow in the system, and that it 
is larger than the limit determined in the previous Section (``spurious flow''), 
below which the method cannot give reliable results. 
In that regime, we compute the statistical error on the integrated flow 
estimate $V_n^\theta\{\infty\}$, and on its average over directions $\theta$, 
due to the limited number of events. 
We shall in particular see that averaging several (typically 4 or 5) values of
$V_n^\theta\{\infty\}$ results in a decrease in the statistical uncertainty 
of a factor of almost 2. 

As explained in Sec.~\ref{s:recipeint}, integrated flow
is determined from the first minimum of
$|G^\theta(ir)|$, which is denoted by $r_0^\theta$.
According to the discussion in Sec.~\ref{s:fluctsample},
for small fluctuations,
this minimum is to leading order the first zero of
$G_{\rm c.l.}(ir)+{\rm Re}\,\delta G^\theta(ir)$.
It differs from $r_0$ [the first zero of $G_{\rm c.l.}(ir)$,
given by Eq.~(\ref{defz0})] by
\begin{equation}
\label{deltar0}
r_0^\theta-r_0= -\frac{{\rm Re}\,\delta G^\theta(ir)}
{\displaystyle\left.
\frac{d G_{\rm c.l.}(ir)}{d r}\right|_{r=r_0}}.
\end{equation}
Using Eq.~(\ref{meanGflow}), and replacing $r_0$ with
its value $j_{01}/V_n$ [Eq.~(\ref{defz0})], one obtains the denominator
of this expression:
\begin{equation}
\label{dGdr}
\frac{d G_{\rm c.l.}(ir)}{d r}|_{r=r_0}=
-V_n\, J_1(j_{01})\,
\exp\!\left(-\frac{j_{01}^2}{4\,\chi^2}\right),
\end{equation}
where $\chi$ is the resolution parameter defined by Eq.~(\ref{defchi}).
Using now Eq.~(\ref{flowestimate0}), and reinserting Eq.~(\ref{dGdr}) in 
Eq.~(\ref{deltar0}), one deduces the statistical fluctuation of the integrated 
flow estimate, $\delta V_n^\theta\equiv V_n^\theta\{\infty\}-V_n$:
\begin{eqnarray}
\label{deltaVn}
\frac{\delta V_n^\theta}{V_n}&=&
-\frac{r_0^\theta-r_0}{r_0}\cr &=&
-\frac{\exp\left(j_{01}^2/(4\,\chi^2)\right)}{j_{01} J_1(j_{01})}\,
{\rm Re}\,\delta G^\theta(i r_0). 
\end{eqnarray}

The correlation $\smean{\delta V_n^\theta \delta V_n^{\theta'}}$
can then be evaluated using Eq.~(\ref{RedeltaG}).
In this equation, the last term in the rhs vanishes since
$r=r'=r_0$, and $G_{\rm c.l.}(ir_0)=0$ by definition of $r_0$.
In order to evaluate the first and second terms in the rhs, we
use the expression of the expectation value $G_{\rm c.l.}(ir)$, 
Eq.~(\ref{meanGflow}), and the two straightforward identities [see
Eq.~(\ref{complex0})]
\begin{eqnarray}
|z+z'|&=&2r_0\cos\left(n(\theta-\theta')/2\right)\cr
|z-z'|&=&2r_0\sin\left(n(\theta-\theta')/2\right).
\end{eqnarray}
One thus obtains
\begin{widetext}
\begin{eqnarray}
\label{thetaprime}
\frac{\mean{\delta V_n^\theta\, \delta V_n^{\theta'}}}{V_n^2}
&=&\frac{1}{2\N\, j_{01}^2\,J_1(j_{01})^2}
\left[
\exp\!\left(\frac{j_{01}^2}{2\chi^2} \cos n(\theta-\theta')\right)
\,J_0\!\left(2\,j_{01}\,\sin\left(\frac{n(\theta-\theta')}{2}\right)
\right)\right. \cr 
& & \qquad\qquad\qquad\qquad\quad +
\left.\exp\!\left(-\frac{j_{01}^2}{2\chi^2} \cos n(\theta-\theta')\right)
\,J_0\!\left(2\,j_{01}
\,\cos\left(\frac{ n(\theta-\theta')}{2}\right)\right)\right].
\end{eqnarray}
Setting $\theta'=\theta$, we obtain the standard deviation on 
$V_n^\theta\{\infty\}$~:
\begin{eqnarray}
\label{stderrtheta}
\frac{\mean{(\delta V_n^\theta)^2}}{V_n^2}
&=&\frac{1}{2\N\, j_{01}^2\,J_1(j_{01})^2}
\left[
\exp\!\left(\frac{j_{01}^2}{2\chi^2}\right)
+
\exp\!\left(-\frac{j_{01}^2}{2\chi^2}\right)
\,J_0\!\left(2\,j_{01}\right)\right].
\end{eqnarray}
\end{widetext}
These relations show that the relative statistical error on integrated flow
depends on the parameters $V_n$ and $\sigma$ only through
the resolution parameter $\chi$,
as explained in Sec.~\ref{s:resolution}.
One sees that the error diverges exponentially for small $\chi$,
while the divergence is only polynomial with finite-order 
cumulants~\cite{Borghini:2001vi}.

\begin{center}
\begin{figure}[ht!]
\centerline{\includegraphics*[width=\linewidth]{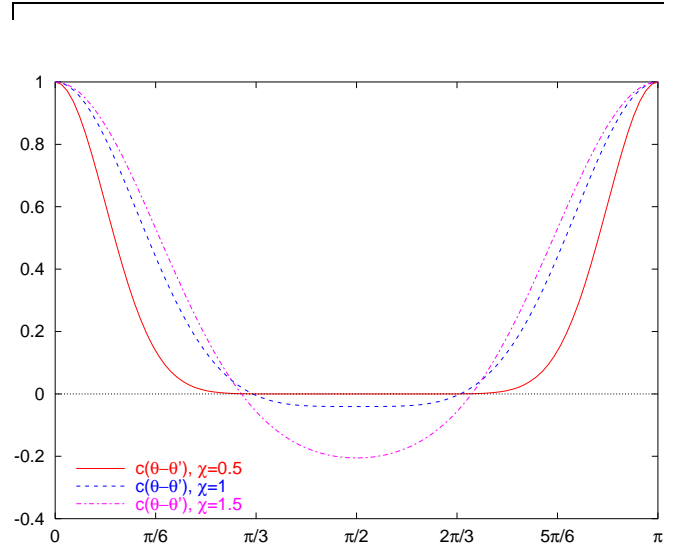}}
\caption{Correlation function $c(\theta-\theta')$, defined by
Eq.~(\ref{ctheta}), as a function of $n(\theta-\theta')$ for
three values of the resolution parameter, $\chi=0.5,1,1.5$.}
\label{fig:correlation}
\end{figure}
\end{center}

The linear correlation coefficient between two estimates $V_n^\theta\{\infty\}$ 
and $V_n^{\theta'}\{\infty\}$ is defined as
\begin{equation}
\label{ctheta}
c(\theta-\theta')\equiv
\frac{\mean{\delta V_n^\theta\, \delta V_n^{\theta'}}}
{\sqrt{\mean{(\delta V_n^\theta)^2}
\mean{(\delta V_n^{\theta'})^2}}}.
\end{equation}
It may vary between $-1$ and 1, which correspond to the cases when
$\delta V_n^{\theta}\{\infty\}$ and $\delta V_n^{\theta'}\{\infty\}$ are
opposite or equal, respectively.
The variation of the correlation coefficient $c(\theta-\theta')$, 
computed with the help of Eqs.~(\ref{thetaprime}) and (\ref{stderrtheta}),
is displayed in Fig.~\ref{fig:correlation} as a function of
the relative angle for several values of $\chi$.
One sees that the correlation is maximum for $\theta'=\theta$
and $\theta'=\theta+\pi/n$, which is natural since
$V_n^{\theta+\pi/n}=V_n^\theta$. For small values of $\chi$,
the correlation vanishes when the relative angle $\theta'-\theta$
is large enough, while for larger values of $\chi$, an
anticorrelation appears around $\theta'=\theta+\pi/2n$.

The shape of the correlation function $c(\theta-\theta')$ shows that 
different estimates (for different values of $\theta$) are not necessarily 
strongly correlated. 
In order to decrease the statistical error, one can thus average 
$V^\theta_n\{\infty\}$ over several values of $\theta$. 
These can be chosen equally spaced, i.e., $\theta=k\pi/p$ 
with $k=0,\cdots,p-1$.
The statistical error on this average
can be derived from Eq.~(\ref{thetaprime}).
Numerical
estimates for various values of $\chi$ and different numbers of values
of $\theta$ are given in Table~\ref{Tint}, where we have assumed that
$\N=20000$ events have been analyzed. 
This Table shows two features. First, the value of the error
badly diverges when the resolution parameter $\chi$ decreases,
so that it is very unlikely that our method can be applied when
$\chi$ is smaller than 0.5, unless a huge number of events are 
available, in agreement with our discussion in Sec.~\ref{s:spurious}. 
Second, the statistical error decreases
when the number of values of $\theta$ increases, but quickly saturates,
so that in practice, 4 or 5 values of $\theta$ are enough.

\begin{table}[ht]
\caption{Relative statistical error on $V_n\{\infty\}$, 
for $\N=20000$ events and various values
of the resolution parameter $\chi$ and the number of
values of $\theta$. }
\label{Tint}
\begin{center}
\begin{tabular}{|c|c|c|c|c|c|}
\hline
Nb of points & $\chi\!=\!0.6$ & $\chi\!=\!0.7$ &
$\chi\!=\!0.8$ & $\chi\!=\!1$ & $\chi\!=\!1.5$ \\
\hline
1& 22.2\%& 7.7\% & 3.8\%& 1.70\%& 0.75\%\\
\hline
2& 15.7\%& 5.4\% & 2.7\%& 1.18\%& 0.46\%\\
\hline
3& 12.8\%& 4.4\% & 2.2\%& 0.98\%& 0.41\%\\
\hline
4& 11.4\%& 4.0\% & 2.1\%& 0.94\%& 0.41\%\\
\hline
5& 11.0\%& 3.9\% & 2.0\%& 0.94\%& 0.41\%\\
\hline
$+\infty$& 10.9\%& 3.9\% & 2.0\%& 0.94\%& 0.41\%\\
\hline
\end{tabular}
\end{center}
\end{table}

The statistical error is of course minimum when the number of values of
$\theta$ goes to infinity.
It is then given by averaging Eq.~(\ref{thetaprime}) over both $\theta$
and $\theta'$. One obtains
\begin{equation}
\frac{\mean{(\delta V_n)^2}}{V_n^2}=
\frac{2}{\N} \frac{1}{j_{01}^2 J_1(j_{01})^2}
\sum_{q=1}^{+\infty} J_q(j_{01})^2 I_q\!\left(\frac{j_{01}^2}{2\chi^2}\right).
\end{equation}
Values in the last line of Table~\ref{Tint} were calculated using this 
expression. 

It is instructive to compare the statistical error on $V_n\{\infty\}$ to 
that on estimates from cumulants of $2k$-particle correlations (where 
$k=1,2,\cdots$), denoted by $V_n\{2k\}$ in Ref.~\cite{Borghini:2001vi}.
In both cases, the statistical error depends on two parameters, namely 
the total number of events, $\N$, and the parameter $\chi$. 
However, the $\chi$-dependence of the errors is very different. 
Thus, whereas the statistical uncertainties on the flow estimates from 
cumulants for small $\chi$ diverges like a power law $1/\chi^{2k}$ 
(see Appendix D in Ref.~\cite{Borghini:2001vi}), the error on $V_n\{\infty\}$ 
depends exponentially on $\chi$ [see Eq.~(\ref{stderrtheta})]. 
That explains why the detected multiplicity per event, which influences the 
value of $\chi$, plays a most crucial role in the present method, while 
cumulants could accommodate lower statistics. 

A numerical comparison is made in Table~\ref{Tint2} for various
values of the resolution parameter $\chi$.
As anticipated, for small $\chi$, the relative error on $V_n\{\infty\}$ is 
significantly larger than that on $V_n\{2k\}$. 
One also notes that the statistical error on $V_n\{\infty\}$ is always 
larger than the error on the standard flow estimate (from two-particle 
correlations) $V_n\{2\}$. 
For values of $\chi$ of order unity, however, the statistical error on 
$V_n\{\infty\}$ is not much larger than the error on the two-particle 
estimate $V_n\{2\}$. 
One must keep in mind here that lower order estimates such as $V_n\{2\}$ 
are biased by nonflow correlations, and that the resulting systematic error 
is usually much larger than the statistical error.

\begin{table}[ht]
\caption{Comparison between different methods: the relative
statistical error on the integrated flow $V_n$ is shown for
the cumulant method~\cite{Borghini:2001vi}, and various
cumulant orders (denoted by $V_n\{2k\}$ with integer $k$),
and for the present method ($V_n\{\infty\}$).
As in the previous table, the number of events is
$\N=20000$ events, and the resolution parameter $\chi$ takes
several values. }
\label{Tint2}
\begin{center}
\begin{tabular}{|c|c|c|c|c|c|}
\hline
 & $\chi\!=\!0.6$ & $\chi\!=\!0.7$ & $\chi\!=\!0.8$ &
$\chi\!=\!1$ & $\chi\!=\!1.5$ \\
\hline
$\delta V_n\{2\}/V_n$ & 1.3\%& 1.0\% & 0.83\%& 0.62\%& 0.37\%\\
\hline
$\delta V_n\{4\}/V_n$ & 4.5\%& 2.7\% & 1.8\%& 1.00\%& 0.43\%\\
\hline
$\delta V_n\{6\}/V_n$ & 7.7\%& 3.7\% & 2.1\%& 0.99\%& 0.41\%\\
\hline
$\delta V_n\{8\}/V_n$ & 9.9\%& 4.1\% & 2.1\%& 0.95\%& 0.41\%\\
\hline
$\delta V_n\{\infty\}/V_n$ & 10.9\%& 3.9\% & 2.0\%& 0.94\%& 0.41\%\\
\hline
\end{tabular}
\end{center}
\end{table}

One also sees clearly in Table~\ref{Tint2} that the error on the present 
estimate $V_n\{\infty\}$ is the limit of the error on $V_n\{2 k\}$ as $k$ 
goes to infinity, for fixed $\chi$.
This reflects the fact that $V_n\{\infty\}$ is itself the limit of $V_n\{2k\}$ 
as $k$ goes to infinity, as explained in Sec.~\ref{s:zeroes}.
It is interesting to note that the statistical error does not increase 
monotonically with the order of the cumulant $2k$, as one might think.
This was already pointed out in Ref.~\cite{Borghini:2001vi}.
For a fixed $\chi$, the error on $V_n\{2k\}$ first increases as $k$ 
increases, reaches a maximum for some value of $k$ and then slightly 
decreases.
This can be seen in the last four columns of Table~\ref{Tint2}.
While the statistical error is always smallest for the standard, two-particle 
estimate $V_n\{2\}$, the error on $V_n\{\infty\}$ is smaller than that on 
the estimate from four-particle correlations $V_n\{4\}$ as soon as $\chi$ 
exceeds $0.89$.
Comparing with the values of $\chi$ given in Table~\ref{Tchi}, this suggests
that the new method can be used without problem at RHIC. The
statistical error on our flow estimate is expected to be comparable 
to that on the estimate from four-particle cumulants, and significantly 
smaller than the systematic error (due to nonflow correlations)
on the estimate from the standard method~\cite{Adler:2002pu}. 

\subsection{Statistical error on the differential flow}
\label{s:statdiff}

We now turn to the statistical uncertainty on the differential flow estimates 
$v^{\prime\theta}_{mn}$ and its average over $\theta$, $v'_{mn}$. 
Note that, unlike the case of integrated flow, the errors we shall compute 
in this Section are {\em absolute} errors, not relative ones. 

Since differential flow is analyzed in principle in a narrow
phase-space window, the statistical error on differential flow
is much larger than the error on integrated flow.
We therefore neglect the statistical error on integrated
flow here:
in Eq.~(\ref{diffflow}), we replace $V_n\{\infty\}$ and
$r_0^\theta$ by their theoretical values, i.e.,
$V_n$ and $r_0$ [see Eq.~(\ref{defz0})].
Similarly, we replace in the denominator the sample
average by an expectation value:
\begin{eqnarray}
\sample{Q^\theta e^{i r_0^\theta Q^\theta}}&\simeq&
\mean{Q^\theta e^{i r_0 Q^\theta}}\cr
&=&-i\,\frac{d G_{\rm c.l.}(ir)}{d r}|_{r=r_0}, 
\end{eqnarray}
where the rhs is given by Eq.~(\ref{dGdr}).
Equation~(\ref{diffflow}) then becomes
\begin{widetext}
\begin{equation}
\label{diffflowf}
v^{\prime\theta}_{mn}\{\infty\}\equiv
(-1)^m\,\frac{e^{j_{01}^2/(4\chi^2)}}{J_m(j_{01})}
\,\sample{\cos\left(mn(\psi-\theta)\right)\,
{\rm Re}\left(i^m G(r_0^\theta\,e^{i\theta})\right)}.
\end{equation}
We now evaluate the correlation between
$v^{\prime\theta}_{mn}\{\infty\}$ and
$v^{\prime\theta'}_{mn}\{\infty\}$, using
an equation similar to Eq.~(\ref{samplingcov}), with 
$\N$ replaced by the total number of ``differential particles'', $\Np$.
We assume, for simplicity, that $\psi$ and $G(z)$ are
statistically independent (which means in particular that
the differential flow vanishes) so that the term involving
$\psi$ factors out. Averaging over $\psi$, this term
gives
\begin{equation}
\mean{\cos(mn(\psi-\theta))\cos(mn(\psi-\theta'))}
=\frac{1}{2}\cos(mn(\theta-\theta')).
\end{equation}
One thus obtains
\begin{equation}
\label{statvprime0}
\mean{\delta v_{mn}^{\prime\theta}\{\infty\}\,
\delta v_{mn}^{\prime\theta'}\{\infty\}}
=\frac{\cos(m n(\theta-\theta'))\,e^{j_{01}^2/(2\chi^2)}}{2\Np\, J_m(j_{01})^2}
\mean{{\rm Re}\,(i^m e^{ir_0Q^\theta})\,
{\rm Re}\,(i^m e^{ir_0 Q^{\theta'}})}.
\end{equation}
Now, ${\rm Re}\,(i^m G(z))$ is, up to a sign, the
real (resp. imaginary) part of $G(z)$ for even
(resp. odd) $m$.
Using Eq.~(\ref{ReImG}) and Eq.~(\ref{meanGflow}), one
finally obtains
\begin{eqnarray}
\label{statvprime1}
\mean{\delta v_{mn}^{\prime\theta}\{\infty\}\,
\delta v_{mn}^{\prime\theta'}\{\infty\}}
&=&\frac{\cos(m n(\theta-\theta'))}{4\Np\, J_m(j_{01})^2}
\left[
\exp\!\left(\frac{j_{01}^2}{2\chi^2} \cos\left
( n(\theta-\theta')\right)\right)
\,J_0\!\left(2j_{01}\,\sin\!\left(\frac{ n(\theta-\theta')}{2}\right)
\right)\right.
\cr & & \qquad\qquad\qquad\qquad +\ 
(-1)^m\left.\exp\!\left(-\frac{j_{01}^2}{2\chi^2} \cos n(\theta-\theta')\right)
\,J_0\!\left(2j_{01}\,\cos\!\left(\frac{ n(\theta-\theta')}{2}\right)
\right)\right],\qquad
\end{eqnarray}
\end{widetext}
where $N'$ is the number of particles in the
considered phase-space bin.
One checks that this expression is invariant under the transformation
$\theta\to\theta+\pi/n$, as expected from the symmetry property
$v_{mn}^{\prime\theta}\{\infty\}=v_{mn}^{\prime\theta+\pi/n}\{\infty\}$.
Setting $\theta=\theta'$ in Eq.~(\ref{statvprime1}) yields the absolute 
statistical error on $v_{mn}^{\prime\theta}$. 

As in the case of integrated flow, one must average over
several values of $\theta$ in order to reduce the statistical
error. Tables~\ref{Tdiff1} and \ref{Tdiff2} show the variation
of the error as the number of values of $\theta$ increases, for the
first two harmonics $v'_n\{\infty\}$ and $v'_{2n}\{\infty\}$.
Once again, 4 or 5 values of $\theta$ are enough in practice to
minimize the statistical error. In the limit when
the number of $\theta$ goes to infinity, the error is
given by
\begin{eqnarray}
\lefteqn{\mean{v'_{mn}\{\infty\}^2}-(v'_{mn})^2=
 \hspace{6cm}} & & \cr
 & & \frac{1}{2\, \Np\,J_m(j_{01})^2}
\sum_{q=-\infty}^{+\infty} J_q(j_{01})^2 
I_{q+m}\!\left(\frac{j_{01}^2}{2\chi^2}\right).
\end{eqnarray}

\begin{table}[ht]
\caption{Statistical error on the reconstructed value
of the differential flow $v'_n$, for $\Np=6\times 10^5$ particles in the bin,
and various values of the resolution parameter and the number of
values of $\theta$.
Note that these are absolute values, not relative values as
in Tables~\ref{Tint} and \ref{Tint2}.}
\label{Tdiff1}
\begin{center}
\begin{tabular}{|c|c|c|c|c|c|}
\hline
Nb of points & $\chi=0.6$ & $\chi=0.7$ & $\chi=0.8$ &
$\chi=1$ & $\chi=1.5$ \\
\hline
1& 6.9\% & 2.4\% & 1.19\% & 0.53\% & 0.24\% \\
\hline
2& 4.9\% & 1.7\% & 0.84\% & 0.37\% & 0.17\% \\
\hline
3& 4.0\% & 1.4\% & 0.69\% & 0.31\% & 0.14\% \\
\hline
4& 3.5\% & 1.2\% & 0.63\% & 0.29\% & 0.14\% \\
\hline
5& 3.4\% & 1.2\% & 0.62\% & 0.29\% & 0.14\% \\
\hline
$+\infty$& 3.3\% & 1.2\% & 0.62\% & 0.29\% & 0.14\% \\
\hline
\end{tabular}
\end{center}
\end{table}

\begin{table}[ht]
\caption{Same as Table~\ref{Tdiff1}, but for the higher harmonic
$v'_{2n}$. }
\label{Tdiff2}
\begin{center}
\begin{tabular}{|c|c|c|c|c|c|}
\hline
Nb of points & $\chi=0.6$ & $\chi=0.7$ & $\chi=0.8$ &
$\chi=1$ & $\chi=1.5$ \\
\hline
1& 8.3\% & 2.9\% & 1.43\% & 0.63\% & 0.28\% \\
\hline
2& 5.9\% & 2.0\% & 1.02\% & 0.46\% & 0.22\% \\
\hline
3& 4.8\% & 1.7\% & 0.83\% & 0.37\% & 0.17\% \\
\hline
4& 4.1\% & 1.4\% & 0.72\% & 0.32\% & 0.15\% \\
\hline
5& 3.8\% & 1.3\% & 0.69\% & 0.32\% & 0.15\% \\
\hline
$+\infty$& 3.7\% & 1.3\% & 0.68\% & 0.32\% & 0.15\% \\
\hline
\end{tabular}
\end{center}
\end{table}

It is also interesting to compare the error computed in this Section 
with the statistical errors on differential flow estimates from multiparticle
cumulants~\cite{Borghini:2001vi}.
This comparison is performed in Tables~\ref{Tdiff3} and \ref{Tdiff4} for the 
first two harmonics $v'_n$ and $v'_{2n}$, respectively.
The same comments apply here as for the integrated flow (Sec.~\ref{s:statint}):
statistical errors on $v'_{mn}\{\infty\}$ are much larger than errors on 
standard flow estimates for small $\chi$, but become comparable, and even 
smaller than estimates from higher-order cumulants, when $\chi$ is close 
to unity or larger.

\begin{table}[ht]
\caption{Comparison between different methods: the
statistical error on the differential flow $v'_n$ is shown for
the estimates from cumulant of 2-particle ($v'_n\{2\}$)
and 4-particle ($v'_n\{4\}$) correlations~\cite{Borghini:2001vi}
and for the present method ($v'_n\{\infty\}$).
As in the previous table,
we assume $\Np=6\times 10^5$ particles in the bin,
the resolution parameter $\chi$ takes several values. }
\label{Tdiff3}
\begin{center}
\begin{tabular}{|c|c|c|c|c|c|}
\hline
 & $\chi=0.6$ & $\chi=0.7$ & $\chi=0.8$ &
$\chi=1$ & $\chi=1.5$ \\
\hline
$\delta v'_n\{2\}$ & 0.18\% & 0.16\%& 0.15\% & 0.13\%& 0.11\%\\
\hline
$\delta v'_n\{4\}$ & 0.88\% & 0.61\%& 0.45\% & 0.29\%& 0.15\%\\
\hline
$\delta v'_n\{\infty\}$ & 3.3\% & 1.2\% & 0.62\% & 0.29\% & 0.14\% \\
\hline
\end{tabular}
\end{center}
\end{table}

\begin{table}[ht]
\caption{Same as Table~\ref{Tdiff4}, but for the higher
harmonic $v'_{2n}$.
Errors on estimates from 3- and 5-particle cumulants~\cite{Borghini:2001vi}
are compared with the present estimate $v'_{2n}\{\infty\}$. }
\label{Tdiff4}
\begin{center}
\begin{tabular}{|c|c|c|c|c|c|}
\hline
 & $\chi=0.6$ & $\chi=0.7$ & $\chi=0.8$ &
$\chi=1$ & $\chi=1.5$ \\
\hline
$\delta v'_{2n}\{3\}$ & 0.48\% & 0.38\%& 0.32\% & 0.24\%& 0.16\%\\
\hline
$\delta v'_{2n}\{5\}$ & 1.42\% & 0.88\%& 0.59\% & 0.33\%& 0.16\%\\
\hline
$\delta v'_{2n}\{\infty\}$ & 3.7\% & 1.3\% & 0.68\% & 0.32\% & 0.15\% \\
\hline
\end{tabular}
\end{center}
\end{table}

In addition, one notes that the error on the higher harmonic $v'_{2n}$ is 
only slightly larger than the error on $v'_n$, while in standard flow 
analyses the error on higher harmonics is often much larger.
It turns out that in the case of higher harmonics, the
statistical error on $v'_{mn}\{\infty\}$ is even smaller than the
statistical error on the estimate from the standard flow
analysis [which we denote by $v'_{mn}\{m+1\}$ since it involves
a ($m+1$)-particle correlation],
if the resolution parameter $\chi$ is large enough.
More precisely, our method yields smaller error bars than the
standard method as soon as $\chi>1.3$ for $v'_{2n}$, $\chi>1.1$ for
$v'_{3n}$, and $\chi>0.94$ for $v'_{4n}$.
Even from the point of view of statistical error bars,
it seems to be the best method to measure higher
harmonics $v_3$, $v_4$, or at least to set upper bounds on their values, 
which could easily be done at SIS or AGS energies where the reaction plane 
resolution is high: correlating more particles may in some cases lead to 
smaller statistical fluctuations, which is a rather unexpected conclusion.
However, one must keep in mind that systematic errors on these higher
harmonics may be large, as shown in Sec.~\ref{s:systematic2}. 

\section{Detector effects}
\label{s:detector}

Until now, we have performed all calculations assuming that the detector 
is perfect, in the sense that it is azimuthally isotropic. 
In this Section, we shall relax this assumption, and investigate the 
influence of acceptance inefficiencies on the determination of flow 
through the present method. 
As we shall see, the generating function will acquire a phase, which does not 
affect the fact that the zeroes lie on the imaginary axis. 
The only sizeable effect on the $\theta$-averaged estimate $V_n\{\infty\}$ 
is of second order in the acceptance coefficients $a_n$ which we soon define, 
and is negligible in most practical cases. 

Let us denote by $A(\phi)$ the probability that a particle with azimuthal 
angle $\phi$ be detected: $A(\phi)$ represents the acceptance-efficiency 
profile of the detector. 
We choose the normalization $\int_0^{2\pi}\!A(\phi)\,d\phi/2\pi=1$.
\footnote{Strictly speaking, such a normalization amounts to assuming that 
the overall efficiency is 100\%, so that it can be larger than 100\% for 
some values of the azimuth if the acceptance is not isotropic\ldots 
However, normalizing $A(\phi)$ {\em a priori} does not influence the 
discussion. 
This simply reflects the fact that the flow analysis is not sensitive to the 
overall efficiency of the detector, but only to the anisotropies in the 
acceptance.}
We assume for simplicity that $A(\phi)$ is independent of the rapidity and 
transverse momentum of the particle. 
Since $A(\phi)$ is a $2\pi$-periodic function, it is natural to expand it in 
Fourier series.
We denote by $a_p$ the corresponding Fourier coefficients:
\begin{equation}
\label{defap}
a_n\equiv \int_0^{2\pi}\!e^{-in\phi}A(\phi)\,\frac{d\phi}{2\pi}.
\end{equation}
The normalization choice of course reads $a_0=1$. 

The probability distribution of $\phi$ for particles seen in the detector, 
for a fixed orientation of the reaction plane $\Phi_R$, is then
\begin{equation}
\label{distriphi-acc}
\frac{dN}{d\phi}=A(\phi)\sum_{n=-\infty}^{+\infty} v_n e^{in(\phi-\Phi_R)},
\end{equation}
where we use the conventions $v_{-n}\equiv v_n$, $v_0\equiv 1$. 

Using this distribution, we can compute the generating function as in
Sec.~\ref{s:Qdistribution}. 
Equation~(\ref{centrallimit}) still holds, but $\mean{Q^\theta|\Phi_R}$ 
is no longer given by Eq.~(\ref{averageQ}), which holds only with a perfect 
acceptance. 
{}From the definition of $Q^\theta$, Eq.~(\ref{defqtheta}), we obtain instead
\begin{equation}
\label{avQdetector}
\mean{Q^\theta|\Phi_R}=\sum_{m=-\infty}^{+\infty}
V_m\,{\rm Re}\left( a_{n-m} e^{i n \theta} e^{-i m \Phi_R}\right),
\end{equation}
which replaces Eq.~(\ref{averageQ}) for an arbitrary detector. 
For a perfect acceptance ($a_0=1$, $a_{n\neq 0}=0$), this expression 
reduces to Eq.~(\ref{averageQ}), as should be. 

We shall now assume for simplicity that only one Fourier harmonic of the 
flow, $v_n$, is non-vanishing, with $n\neq 0$. 
This is most probably a reasonable assumption at ultrarelativistic energies. 
Then, only three terms contribute in the sum, $m=-n$, $m=0$ and $m=n$. 
We may then rewrite the previous equation as 
\begin{eqnarray}
\label{avQdetector1}
\mean{Q^\theta|\Phi_R} & = & V_0\,{\rm Re}\left( a_n e^{in\theta} \right)\cr
 & & +\ V_n\,{\rm Re}\!\left[ 
\left(1 + a_{2n} e^{2in\theta}\right)\,e^{in(\Phi_R-\theta)} \right], \qquad
\end{eqnarray}
where $V_0=\smean{\sum w_j}$ and we have used the normalization condition 
$a_0=1$, as well as the properties $a_{-2n}=a_{2n}^*$, see Eq.~(\ref{defap}), 
and $V_{-n}=V_n$, which follows from Eqs.~(\ref{flowvector}) and (\ref{defVn}). 
Inserting this value in Eq.~(\ref{centrallimit}), we obtain 
\begin{widetext}
\begin{equation}
\label{GphiRdetector}
\mean{e^{zQ^\theta}|\Phi_R}=
\exp\left(z\,V_0\,{\rm Re}\left( a_n e^{in\theta} \right) + 
 \frac{\sigma^2 z^2}{4} + 
 z\,V_n\,{\rm Re}\!\left[ \left(1+a_{2n} e^{2in\theta}\right) 
   e^{in (\Phi_R-\theta)}\right] \right).
\end{equation}
\end{widetext}
Averaging this expression over $\Phi_R$ (we still assume that $\sigma$ is 
independent of $\Phi_R$), we obtain
\begin{equation}
\label{Gtheta_acc}
G^\theta(z)=
e^{z\,V_0\,{\rm Re}( a_n e^{in\theta}) + \sigma^2 z^2/4} 
\,I_0\!\left(z\,V_n\left|1 + a_{2n} e^{2in\theta}\right|\right).
\end{equation}
Comparing this result with the perfect-acceptance case in 
Eq.~(\ref{meanGflowbis}), there appear several modifications.
First of all, $G^\theta(z)$ is no longer real-valued for values of $z$ on 
the imaginary axis. 
Then, there is a new term in the exponential factor, which involves the 
acceptance coefficient $a_n$. 
This term cannot produce any spurious zero of $G^\theta(z)$, nor shift the 
position of the zeroes. 
When $z$ is purely imaginary, it is a pure phase, which goes away when 
taking the modulus $|G^\theta(z)|$. 
However, this phase makes the real part of $G^\theta(z)$ oscillate, and 
thus makes it vanish on the imaginary axis. 
This explains why it was important to consider (the minimum of) the modulus, 
not (the zero of) the real part, as mentioned in Sec.~\ref{s:Qdistribution}. 

The remaining modification is more relevant for the procedure introduced in 
Sec.~\ref{s:recipeint}, since it results in a shifting of the position of the 
generating-function zeroes. 
The argument of the Bessel function in Eq.~(\ref{Gtheta_acc}) now depends 
on $\theta$, and involves the acceptance coefficient $a_{2n}$. 
Thus, using Eq.~(\ref{flowestimate0}), we now obtain
\begin{equation}
\label{vintdetector}
V_n^\theta\{\infty\}=V_n \left|1 +a_{2n} e^{2in\theta}\right|, 
\end{equation}
where $a_{2n}$ is given by Eq.~(\ref{defap}): the estimate $V_n^\theta$ 
now depends on $\theta$, as can be seen in Fig.~\ref{fig:vacc}.
As expected from the general discussion in Sec.~\ref{s:recipeint}, 
it is ($\pi/n$)-periodic.
In the case of a detector with a hole of $\alpha$ radians around $\phi=\pi$, 
an explicit calculation gives $a_{2n}=-\sin(n\alpha)/[(2\pi-\alpha)n]$. 
Thus, in the simulation of Sec.~\ref{s:recipedetector}, $a_4\simeq 0.0827$, 
quite a small number, but large enough to yield the $\theta$-dependence 
observed in Fig.~\ref{fig:vacc}.

Averaging Eq.~(\ref{vintdetector}) over $\theta$, one obtains 
\begin{equation}
\label{vintdetectormean}
V_n\{\infty\}=V_n (1+|a_{2n}|^2). 
\end{equation}
The relative error is $|a_{2n}|^2$, which is in practice small
if the detector has a reasonable azimuthal coverage. 

Consider now differential flow. 
We leave open the possibility that the corresponding acceptance function 
is not the same as for integrated flow, and we denote by $a'_n$ the 
Fourier acceptance coefficients for the differential particle. 
For the sake of brevity, we shall only consider the case $m=1$ (differential 
flow in the same harmonic as integrated flow); the generalization to 
$m$ arbitrary only involves much more tedious calculations. 
Finally, we make the same assumption as above, namely, that only one flow 
harmonic does not vanish. 

Under this assumption, Eq.~(\ref{average1}) is replaced with an equation 
similar to Eq.~(\ref{avQdetector1})
\begin{equation}
\label{avpsidetector1}
\mean{\cos(n(\psi-\theta))|\Phi_R}=
v'_n\,{\rm Re}\left[ (1+a'_{2n} e^{2in\theta}) e^{in(\Phi_R-\theta)}\right]. 
\end{equation}

This identity can be then combined with Eq.~(\ref{GphiRdetector}) to compute
$D_m^\theta(z)$ defined by Eq.~(\ref{defD(z)}).
To perform the integration over $\Phi_R$, one can use the following integral: 
\begin{equation}
\int_0^{2\pi}\!
{\rm Re}\left(B e^{im\varphi}\right) e^{z\,{\rm Re}(Ae^{i\varphi})}\,
\frac{d\varphi}{2\pi} =
\frac{{\rm Re}(B\,{A^*}^m)}{|A|^m}\,I_m(z|A|), 
\end{equation}
valid for arbitrary complex numbers $A$ and $B$. 
We use this result with $m=1$ (other values of $m$ might be useful for 
higher harmonics) and $\varphi=\Phi_R-\theta$:
\begin{widetext}
\begin{equation}
D_1^\theta(z) = e^{z\,V_0\,{\rm Re}( a_n e^{in\theta}) + \sigma^2 z^2/4}\,
\frac{{\rm Re}\left[(1+a'_{2n} e^{2in\theta})(1+a_{2n}^* e^{-2in\theta})\right]}
{\left|1+a_{2n} e^{2in\theta}\right|}\,
I_1\!\left(zV_n\left|1+a_{2n} e^{2in\theta}\right|\right)\,v'_n.
\end{equation}
\end{widetext}
Note that the exponential prefactor and the argument of the Bessel function 
are the same as in Eq.~(\ref{Gtheta_acc}). 
After some algebra, similar to that in Sec.~\ref{s:asympdiff}, one finally 
obtains
\begin{equation}
\label{v'theta-acc}
v^{\prime\theta}_n\{\infty\}= v'_n\,
\frac{{\rm Re}\left[(1+a'_{2n} e^{2in\theta})(1+a_{2n}^* e^{-2in\theta})\right]}
{\left|1+a_{2n} e^{2in\theta}\right|}.
\end{equation}
As in the case of integrated flow, acceptance inefficiencies result in the 
appearance of a $\theta$-dependent, ($\pi/n$)-periodic, 
multiplicative coefficient in front of $v'_n$. 
In the particular case where $a'_{2n}=a_{2n}$, Eq.~(\ref{v'theta-acc}) gives 
\begin{equation}
v^{\prime\theta}_n\{\infty\}= v'_n \left|1 +a_{2n} e^{2i n \theta}\right|.
\end{equation}
Integrating over phase space, one recovers Eq.~(\ref{vintdetector}).

\section{Concluding remarks}
\label{s:summary}

In this paper, we have introduced a new method of analysis of anisotropic 
flow in heavy-ion collisions. 
This is the first method to extract both integrated and differential flow 
from the interparticle correlations between a large number of particles, 
rather than from the correlations between only a small number. 
Thus, it is able to isolate a collective effect which involves all 
particles, such as flow, removing the influence of other, ``nonflow'' effects, 
which can bias few-particle methods. 
The flow estimates obtained with the new method are therefore more reliable 
than with any other method. 

This increased reliability is of course important inasmuch as flow is 
concerned: with more accurate collective flow estimates, the constraints on 
models become stronger. 
But it also matters for the measurement of other effects, as for instance 
when reconstructing jets from the azimuthal correlations they induce between 
high momentum particles at ultrarelativistic energies~\cite{Adler:2002ct,%
  Adler:2002tq,Agakichiev:2003gg}. 
At high $p_T$, (elliptic) flow is large, and constitutes a huge background. 
It is most important to know its magnitude accurately to extract nonflow
correlations.

Let us enter into a little more detail. 
The method relies on the search of the first minimum of a generating 
function of multiparticle azimuthal correlations; the actual position of 
this minimum directly yields an integrated flow estimate. 
Once the minimum has been found, computing a second function at its 
position gives differential flow. 
This procedure is simpler to implement than other methods of flow analysis. 
It is significantly faster numerically, and less tedious, than the methods
based on cumulants of multiparticle correlations. 
In addition, in the differential flow analysis there is no need to subtract 
autocorrelations, which are negligible. 
Finally, detector effects are negligible as well in most cases. 
We have carefully studied the various sources of errors, either systematic 
errors intrinsic to the method (due to nonflow effects, higher flow harmonics, 
detector inefficiencies), or statistical errors. 
This allows us to conclude that it is the most accurate method to measure 
both directed and elliptic flow in collisions from 100~MeV to a few GeV per 
nucleon (SIS and AGS energies) and elliptic flow at ultrarelativistic 
energies, especially at RHIC and the CERN Large Hadron Collider (LHC). 

It may be possible to extend the method, so as to be able to measure 
directed flow at ultrarelativistic energies as well, paralleling the step
which led from the cumulant method of Refs.~\cite{Borghini:2000sa,%
  Borghini:2001vi} to that of Ref.~\cite{Borghini:2002vp}. 
Namely, the generating function of one complex variable $G(z)$ could 
be generalized to a two-variable function $G(z_1, z_2)$ involving azimuthal 
correlations in two different harmonics (corresponding to $v_1$ and $v_2$) 
at once. 
This new function could then be used to measure directed flow using as 
a reference the elliptic flow, which is large at ultrarelativistic energies, 
and thus provides a good reference. 

Another generalization of the method would be to extract directly nonflow 
azimuthal correlations, and especially, to obtain them with respect to the 
impact-parameter direction in non-central collisions. 
This could be done, for instance, to study the azimuthal dependence of 
Hanbury--Brown Twiss (HBT) correlations~\cite{Lisa:2000xj}, or that of jet 
quenching~\cite{Adler:2002tq}. 

Finally, thanks to the generality of its formalism, this method could easily 
be extended to other types of observables where ``fluctuations'' are of 
interest.
It seems to be the most natural method to look for critical, large scale 
fluctuations in a system, which are expected in the vicinity of 
a phase transition~\cite{Stephanov:1999zu}.

\section*{Acknowledgments}

J.-Y.~O. thanks B. Duplantier and J.~Zinn-Justin for discussions, and 
E.~A.~De~Wolf for providing us with a copy of Ref.~\cite{dewolf}.
R.~S.~B. acknowledges the hospitality of the SPhT, CEA, Saclay;
J.-Y.~O. acknowledges the hospitality of the Department of Theoretical
Physics, TIFR, Mumbai.
Both acknowledge the financial support from CEFIPRA, New Delhi, 
under its project no. 2104-02.
N.~B. acknowledges support of the Bergen Computational Physics Laboratory in
the framework of the European Community --- Access to Research Infrastructure
action of the Improving Human Potential Programme.

\appendix

\section{An alternative form of the generating function}
\label{s:Gtilde}

The generating function on which our method is based, $G^\theta(z)$, 
has been defined in Eqs.~(\ref{defqtheta}) and (\ref{defGint}):
\begin{equation}
G^\theta(z)=\sample{\exp\!\left(z\sum_{j=1}^M w_j\cos(n(\phi_j-\theta))
\right)}.
\end{equation}
An alternative choice is:
\begin{equation}
\label{defGtilde}
\tilde G^\theta(z) = 
\sample{\prod_{j=1}^M\left[ 1+z w_j\cos(n(\phi_j-\theta)) \right]}.
\end{equation}
The procedure to determine the integrated flow with $\tilde G^\theta(z)$
is exactly the same as for $G^\theta(z)$, as presented in
Sec.~\ref{s:recipeint}.
In order to determine the differential flow, one should 
replace $e^{i r_0^\theta Q^\theta}$ in the numerator of the 
rhs of Eq.~(\ref{diffflow}) with the term in curly brackets
in the rhs of Eq.~(\ref{defGtilde}), with $z=ir_0^\theta$;
next, one should replace $\sample{Q^\theta e^{i r_0^\theta Q^\theta}}$ 
in the denominator of the rhs of Eq.~(\ref{diffflow}) with 
the derivative $d\tilde G^\theta/dz$, evaluated at $z=ir_0^\theta$. 

The essential point is that $\tilde G^\theta(z)$ factorizes for
independent subsystems, as does $G^\theta(z)$, so that the discussion
in Secs.~\ref{s:cumulants} and \ref{s:Qdistribution} 
can be readily extended to $\tilde G^\theta(z)$.
The only difference is that $\sigma$ in Eq.~(\ref{centrallimit}) 
is no longer given by Eq.~(\ref{defsigma2}). 
In fact, $\sigma$ vanishes for independent particles, and is 
generally smaller for $\tilde G^\theta$ than for $G^\theta$. 
Although the value of the generating function is modified, the zeroes 
remain at the same position, and in particular the first one, from which 
flow (either integrated or differential) is determined.
A similar equivalence is also known in statistical 
mechanics~\cite{nienhuis}.

A generating function similar to $\tilde G^\theta(z)$ was chosen
in the cumulant expansion of Ref.~\cite{Borghini:2001vi}.
The argument was that when expanding in powers of $z$,
$\tilde G^\theta(z)$ involves only correlations between different
particles, while $G^\theta(z)$ involves ``autocorrelation'' terms.
The bias induced by these autocorrelations was studied in detail in
Ref.~\cite{Borghini:2000sa} for finite-order cumulants. 
In general, this bias is of the same order of magnitude 
as the bias induced by nonflow correlations, as
explained in Sec.~\ref{s:autocorrelations}. 

Let us illustrate this by a simple explicit example. 
We repeat the calculations of Sec.~\ref{s:sensitivity} with 
$\tilde G^\theta(z)$ instead of $G^\theta(z)$:
particles are emitted in collinear jets containing $q$ particles 
each. If the jet angles are randomly distributed, one obtains
\begin{equation}
\tilde G^\theta(z)=\mean{(1+z\cos\phi)^q}^{M/q},
\end{equation}
where angular brackets denote an average over $\phi$, which gives:
\begin{equation}
\tilde G^\theta(z)=\left(\sum_{l=0}^{[q/2]}\frac{z^{2l}}{(q-2l)!
\,(l!)^2\, 2^{2l}}\right)^{M/q}.
\end{equation}
For independent particles ($q=1$) $\tilde G^\theta(z)=1$ 
[compare with Eq.~(\ref{Gnoflow+jets})]. 
This function has no zero, hence the analysis yields no spurious flow.
This shows that $\tilde G^\theta(z)$ is an improvement over 
$G^\theta(z)$ when there is no nonflow correlation. 
For higher values of $q$, $\tilde G^\theta(z)$ is 
a polynomial whose roots can be computed numerically. 
The estimate $V_n\{\infty\}$ is then given by
Eq.~(\ref{flowestimate0}). 
One obtains the values $1.70$, $2.94$, $4.07$ for $q=2,3,4$ 
respectively. This is very close to the value $V_n\{\infty\}=q$
obtained with $G^\theta(z)$ [see Eq.~(\ref{spuriousinfty})], 
which shows that $\tilde G^\theta(z)$ no longer represents 
an improvement over $G^\theta(z)$ when nonflow correlations 
are present. 

The same conclusions hold for systematic errors, whose order 
of magnitude is not modified by autocorrelations. 
In practice, the latter may increase the contribution of 
each term in the rhs of Eq.~(\ref{systerror}), but 
this equation remains valid as an order of magnitude, 
with either form of the generating function.

The final results of Secs.~\ref{s:systematic} to 
\ref{s:detector} also apply when $\tilde G^\theta(z)$ is used 
instead of $G^\theta(z)$, although intermediate calculations may
differ. 
In particular, a derivation of statistical errors with a 
generating function similar  to $\tilde G^\theta(z)$ is given 
in Appendix D of \cite{Borghini:2001vi}. 

As a conclusion, $\tilde G^\theta(z)$ is expected to give more
accurate results than $G^\theta(z)$ when nonflow correlations are weak. 
The price to pay is that $\tilde G^\theta(z)$ 
requires much more computer time:
for each value of $z$, one needs to evaluate the product over all
particles in Eq.~(\ref{defGtilde}), instead of calculating only one
flow vector, valid for all values of $z$.
Furthermore, since the motivation of this paper was to obtain a 
method which remains valid when nonflow correlations are present, 
and since $\tilde G^\theta(z)$ is not better than $G^\theta(z)$ in 
this respect, we chose to use $G^\theta(z)$ in the paper. 

\section{Relation with the cumulant expansion of
        Ref.~\cite{Borghini:2000sa}}
\label{s:equivalence}

In Ref.~\cite{Borghini:2000sa}, integrated flow was determined from
the following generating function (Eq.~(B2) of Ref.~\cite{Borghini:2000sa}):
\begin{equation}
G(z_1,z_2)=\sample{e^{z_1 Q + z_2 Q^*}},
\end{equation}
where $z_1$ and $z_2$ are two independent complex variables,
$Q\equiv Q_x+iQ_y$, and $Q^*=Q_x-iQ_y$.
Cumulants $\cumul{Q^k (Q^{*})^l}$ were then defined by the
power-series expansion of $\ln G(z_1,z_2)$:
\begin{equation}
\label{oldcumulants}
\ln G(z_1,z_2)=\sum_{k,l}\frac{z_1^kz_2^l}{k!\,l!}\cumul{Q^k (Q^{*})^l}. 
\end{equation}
The new generating function, Eq.~(\ref{defGint}), can be expressed in terms
of $G(z_1,z_2)$:
\begin{equation}
G^\theta(z)=G\!\left(\frac{z e^{-i\theta}}{2},\frac{z e^{i\theta}}{2}\right).
\end{equation}
The cumulant $\cumul{(Q^\theta)^k}$ defined by Eq.~(\ref{defcumulint})
is a linear combination of the $\cumul{Q^{k-l} (Q^{*})^l}$:
\begin{equation}
\label{newvsold}
\cumul{(Q^\theta)^k}=\frac{1}{2^k}\sum_{l=0}^k
\left(\!\begin{array}{c} k \\ l \end{array}\!\right)
e^{i(2l-k)\theta}\cumul{Q^{k-l} (Q^{*})^l}.
\end{equation}
In Ref.~\cite{Borghini:2000sa}, the recommended procedure was to
extract an estimate of the flow, denoted by
$V_n\{2k\}$ in Ref.~\cite{Borghini:2001vi},
using the ``diagonal'' cumulants $\cumul{Q^k (Q^*)^k}$,
which are the only non-vanishing cumulants
except for statistical fluctuations and detector asymmetries.
By definition of the estimates, $\cumul{Q^k (Q^*)^k}$ is equal to
$(V_n\{2k\})^{2k}$, up to a multiplicative factor.

In the present paper, we obtain an estimate $V_n$ from
$\cumul{(Q^\theta)^{2k}}$, which was denoted by
$V_n^{\theta}\{2 k\}$ in Sec.~\ref{s:Qdistribution}:
By definition, $\cumul{(Q^\theta)^{2k}}$ is equal to
$(V_n^\theta\{2k\})^{2k}$, up to a multiplicative factor.
In order to obtain the relation between $V_n^{\theta}\{2k\}$ and
$V_n\{2k\}$, let us average Eq.~(\ref{newvsold}) over $\theta$:
\begin{equation}
\label{newvsold1}
\int_{0}^{2\pi}\frac{d\theta}{2\pi}\cumul{(Q^\theta)^{2k}} = 
\frac{(2k)!}{2^{2k}(k!)^2}\cumul{Q^k (Q^*)^k}.
\end{equation}
One then obtains the following relation:
\begin{equation}
\int_{0}^{2\pi}\frac{d\theta}{2\pi} (V_n^\theta\{2k\})^{2k} = 
(V_n\{2k\})^{2k}.
\end{equation}
In the limit of an infinite number of events, and with a perfectly
symmetric detector $V_n^\theta\{2k\}$ is independent of $\theta$
by azimuthal symmetry. The above equation shows that
$V_n^\theta\{2k\}$ then coincides exactly with the estimate $V_n\{2k\}$.
In this case, our estimate $V_n\{\infty\}$ is exactly
the large-order limit of $V_n\{2k\}$ for large $k$.

However, this is no longer the case when statistical fluctuations and
detector asymmetries are taken into account. In this paper,
we first let the order $k$ go to infinity for a fixed value of
$\theta$, and obtain the limit $V_n^\theta\{\infty\}$. Then,
we average over $\theta$. On the other hand, 
in the method of Ref.~\cite{Borghini:2000sa}, one first averages
over $\theta$ for a fixed $k$, which is natural when working
at a finite cumulant order.

This is actually a minor difference. The only practical consequence
is that acceptance corrections, which must be applied when the detector
has limited azimuthal coverage, differ.
The present approach, where the whole analysis is
done for a fixed value of $\theta$, is more natural and leads
to simpler acceptance corrections (see Sec.~\ref{s:detector})
than the earlier method (for which the acceptance corrections
were derived in Appendix~C of Ref.~\cite{Borghini:2001vi}).

Another difference with the method of~\cite{Borghini:2000sa} is 
that, in the latter,  weights $w_j$ in Eq.~(\ref{flowvector})  were chosen 
proportional to $1/\sqrt{M}$, while they are independent of 
$M$ in the present paper. The rationale was that lower-order cumulants 
contain large trivial contributions from ``autocorrelations'': 
for instance, $\mean{|Q^2|}$ is nonvanishing even if there is no
flow. With the chosen weight, these contributions were independent 
of $M$ and could easily be subtracted. However, the magnitude 
of these autocorrelations becomes smaller and smaller as the 
cumulant order increases, and they are negligible with the present 
method (which is the limit of asymptotically large order), so that 
$1/\sqrt{M}$ factors are no longer required.

In Ref.~\cite{Borghini:2001vi}, an improvement over the method 
of Ref.~\cite{Borghini:2000sa} was proposed, using a generating 
function similar to $\tilde G^\theta(z)$ in Eq.~(\ref{defGtilde}). 
This new formulation was free from autocorrelations. 
However, we pointed out that when the detector has limited azimuthal 
coverage, multiplicity fluctuations induce errors on lower-order
cumulants which are similar to the errors induced by autocorrelations.
These errors were minimized by taking weights proportional to
$1/M$, and using a slightly modified definition of the generating 
function of cumulants (Eq.~(7) of Ref.~\cite{Borghini:2001vi}).  
With the present method, such a refinement is not necessary.

In this paper, we choose to work with weights which are independent 
of $M$ for two reasons:
first, this simplifies the discussion of Sec.~\ref{s:zeroes} and 
makes the relation to Lee-Yang theory clearer. Second, the 
multiplicity may not be a relevant quantity in heavy-ion 
experiments at lower energies, where a nuclear fragment 
contributes as much to the flow as several nucleons. 
However, our analysis could as well be performed with 
$M$-dependent weights. The only difference is that the
integrated flow $V_n$ would scale differently with the 
multiplicity, and its fluctuations with impact parameter, 
studied in Sec.~\ref{s:fluctuations}, would also differ.

\section{Large orders}
\label{s:largeorders}

\begin{center}
\begin{figure}[ht!]
\centerline{\includegraphics*[width=0.9\linewidth]{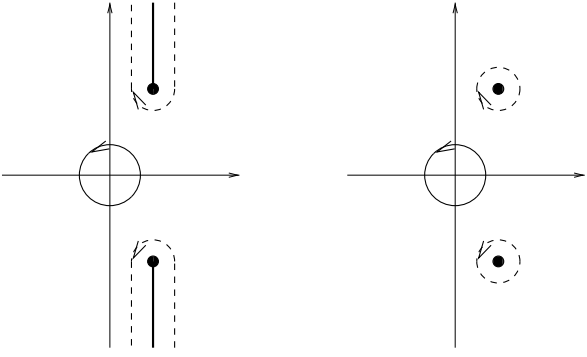}}
\caption{Integration contours in the case where $f(z)$
has a logarithmic singularity (left) or a pole singularity (right).
The full line is the initial integration contour,
and the dashed line is the deformed contour.}
\label{fig:cauchy}
\end{figure}
\end{center}

Consider a function $f(z)$ which is analytic in the vicinity of the origin. 
It can be expanded in power series:
\begin{equation}
f(z)=\sum_{k=0}^{\infty} a_k z^k.
\end{equation}
The coefficients $a_k$ can be expressed in integral form:
\begin{equation}
\label{cauchy}
a_k=\oint \frac{f(z)}{z^{k+1}}\frac{dz}{2 i\pi},
\end{equation}
where the integration contour circles the origin in the complex
plane (full line in Fig.~\ref{fig:cauchy}).
The integration contour can then be expanded until it
reaches a singularity of $f(z)$ (see Fig.~\ref{fig:cauchy}).
We are interested in the asymptotic behavior of
the coefficients $a_k$ as $k$ goes to infinity.
For large $k$, the integral in Eq.~(\ref{cauchy}) is
dominated by the smallest values of $|z|$.
Therefore, the large-order behavior of $a_k$ is determined
by the singularities of $f(z)$ which are closest to the origin.

The first case of interest in this paper (Sec.~\ref{s:cumulants})
is $f(z)=\ln G^\theta(z)$, where $G^\theta(z)$
is analytic in the whole complex plane (entire),
and satisfies the symmetry relation Eq.~(\ref{symmetry2})
(i.e., it is real for real $z$).
Let us denote by $z_0$ the zero of $G(z)$ which is closest
to the origin in the upper plane.
We assume that it is a simple zero.
Thanks to Eq.~(\ref{symmetry2}), its complex conjugate $z_0^*$ is also a zero.
The singularity of $f(z)$ at these points is logarithmic,
and the discontinuity of $f(z)$ across the cut
(Fig.~\ref{fig:cauchy}, left) is $-2i\pi$.
The integral in Eq.~(\ref{cauchy}) then reduces to
\begin{eqnarray}
\label{logcauchy}
a_k &\sim&-\int_{z_0}^{\infty} \frac{dz}{z^{k+1}}
-\int_{z_0^*}^{\infty} \frac{dz}{z^{k+1}}\cr
&=&-\frac{2}{k}{\rm Re}\left(\frac{1}{(z_0)^k}\right).
\end{eqnarray}
It is real, as expected since $G^\theta(z)$ is real for real $z$.

The second case we are interested in (see Sec.~\ref{s:asympdiff})
is $f(z)=D^\theta_m(z)/G^\theta(z)$, where $G^\theta(z)$
is the same function as above,
and $D^\theta_m(z)$ is another function sharing the same properties:
$D^\theta_m(z)$ is entire and also satisfies the symmetry relation
Eq.~(\ref{symmetryd2}). 
The singularities of $f(z)$ which are closest to the origin
are $z_0$ and $z_0^*$, as in the previous case.
They are now {\em pole} singularities, and their contribution
to the integral in Eq.~(\ref{cauchy}) is given by the
theorem of residues:
\begin{eqnarray}
\label{polecauchy}
a_k &\sim&-\frac{D^\theta_m(z_0)}{(z_0)^{k+1}(G^\theta)'(z_0)}
-\frac{D^\theta_m(z_0^*)}{(z_0^*)^{k+1}(G^\theta)'(z_0^*)}\cr
&=&-{\rm Re}\left(\frac{2}{(z_0)^{k+1}} \,
\frac{D^\theta_m(z_0)}{
(G^\theta)'(z_0)}\right),
\end{eqnarray}
where $(G^\theta)'$ denotes the derivative of $G^\theta(z)$
with respect to $z$.

\section{A quantitative estimate of systematic errors}
\label{s:systerrors}

Because of flow itself, and even in the case of a perfect detector, the 
standard deviation $\sigma$ defined in Eq.~(\ref{defsigma2}) 
depends on the orientation of the reaction plane $\Phi_R$.
In this Appendix, we shall explicitly compute the systematic error on 
integrated and differential flows due to this dependence, assuming 
that particles are independent (no nonflow correlations). 
In addition, we shall show that this dependence is influenced by the higher 
flow harmonic $v_{2n}$, which thus contaminates the measurement of $v_n$. 
Throughout the Appendix, we assume that the detector has perfect azimuthal 
isotropy. 

For a fixed orientation of the reaction plane $\Phi_R$, the 
normalized azimuthal distribution is 
\begin{equation}
\label{distriphi}
\frac{dN}{d\phi}=\frac{1}{2\pi}\sum_{n=-\infty}^{+\infty}
v_n\cos\left[n(\phi-\Phi_R)\right],
\end{equation}
where we define $v_0\equiv 1$ and $v_{-n}=v_n$. 

In order to compute the generating function (\ref{defGint}) 
in various physical situations, it is useful to first
evaluate the following single-particle average
with the azimuthal distribution (\ref{distriphi}): 
\begin{eqnarray}
\label{nf2}
\mean{e^{z\cos(n(\phi-\theta))}|\Phi_R} &\!\!=\!\!&
\int_0^{2\pi}\!\frac{dN}{d\phi}\,e^{z\cos(n(\phi-\theta))} d\phi\cr
&\!\!=\!\!&\sum_{m=-\infty}^{+\infty}\! v_{mn} I_m(z)\,\cos(mn(\Phi_R-\theta)). 
\cr
& & 
\end{eqnarray}
Please note that both $m$ and $-m$ contribute equally to the sum. 

Let us now consider a simple model of the collision: we assume 
that the azimuthal angles of the particles are independent
for a fixed orientation of the reaction plane $\Phi_R$, i.e., 
we assume that all azimuthal correlations are due to flow.
If $Q^\theta$ is defined with unit weights in Eq.~(\ref{defqtheta}), 
then 
\begin{equation}
\mean{e^{zQ^\theta}|\Phi_R}=\mean{e^{z\cos(n(\phi-\theta))}|\Phi_R}^M. 
\end{equation}
Taking the logarithm of this expression, using Eq.~(\ref{nf2}) and expanding 
in powers of $z$ up to order $z^2$, we obtain
\begin{eqnarray}
\label{shortrangeapp}
\ln\mean{e^{zQ^\theta}|\Phi_R}&=&
M v_n \cos(n(\Phi_R-\theta))\,z \cr
 & & \ -\ M\left(v_{2n}-v_n^2\right)\sin^2(n(\Phi_R-\theta))\frac{z^2}{2} \cr
 & & \ +\ M\left(1+v_{2n}-2v_n^2\right) \frac{z^2}{4}.
\end{eqnarray}
Comparing this result with Eq.~(\ref{centrallimit}), one sees that the 
standard deviation $\sigma$ depends on $\Phi_R$. 
This dependence was neglected in Sec.~\ref{s:Qdistribution}. 
We shall now assume that it is a small correction, and evaluate the 
resulting systematic error on our flow estimates. 

For that purpose, we introduce the following notations:
\begin{eqnarray}
\label{abreviations}
Z&\equiv& Mv_n z\cr
\alpha &\equiv& n(\Phi_R-\theta)\cr
\epsilon&\equiv&\frac{v_{2n}-v_n^2}{2 M v_n^2}.
\end{eqnarray}
With these notations, we rewrite Eq.~(\ref{shortrangeapp}) as 
\begin{equation}
\label{shortrangeapp1}
\mean{e^{zQ^\theta}|\Phi_R}=e^{M(1+v_{2n}-2v_n^2)z^2/4}\,
e^{Z\cos\alpha -\epsilon Z^2\sin^2\alpha}.
\end{equation}
In order to compute the various generating functions, we 
shall need to evaluate the following integral:
\begin{widetext}
\begin{equation}
\int_0^{2\pi}\!\cos(m\alpha)\, e^{Z\cos\alpha}\, Z^2\sin^2\alpha\,
\frac{d\alpha}{2\pi}=
-\int_0^{2\pi}\!\cos(m\alpha)\, e^{Z\cos\alpha}\left(m^2-Z\cos\alpha\right)
\frac{d\alpha}{2\pi},
\end{equation}
which is obtained with two successive integrations by parts. 
Using this identity, one derives the following result, which is valid to 
first order in the small parameter $\epsilon$:
\begin{eqnarray}
\label{lemme}
\int_0^{2\pi}\!\cos(m\alpha)\, e^{Z\cos\alpha-\epsilon Z^2\sin^2\alpha}\,
\frac{d\alpha}{2\pi}
 & = & (1+\epsilon m^2) \int_0^{2\pi}\!
\cos(m\alpha)\, e^{Z(1-\epsilon)\cos\alpha}\, \frac{d\alpha}{2\pi}\cr
 & = & (1+\epsilon m^2)\, I_m(Z(1-\epsilon)). 
\end{eqnarray}
\end{widetext}
Using Eqs.~(\ref{shortrangeapp1}) and (\ref{lemme}) with $m=0$, we obtain 
\begin{eqnarray}
G^\theta(z) &=& 
\int_0^{2\pi}\!\mean{e^{zQ^\theta}|\Phi_R} \frac{d\Phi_R}{2\pi}\cr
&=& e^{M(1+v_{2n}-2v_n^2)z^2/4}\,I_0(Mv_n z(1-\epsilon)). \qquad
\end{eqnarray}
The first zero of $G^\theta(z)$ on the upper half of the imaginary axis 
thus lies at 
\begin{equation}
\label{z0syst}
z^\theta_0=ir^\theta_0=\frac{ij_{01}}{Mv_n (1-\epsilon)}.
\end{equation}
The corresponding estimate of the integrated flow is then defined by 
Eq.~(\ref{flowestimate0}):
\begin{equation}
\label{vnappsyst}
V_n\{\infty\}=V_n (1-\epsilon),
\end{equation}
where $V_n=Mv_n$ is the exact result. The relative 
error on the integrated flow is $-\epsilon$. Its expression, 
Eq.~(\ref{abreviations}), is 
in agreement with the expected order of magnitude, Eq.~(\ref{systerror}). 
[Remember that the second term in that equation, in $1/(Mv_n)^2$, is due 
to the term in $z^3$ in the expansion of $\ln\smean{e^{zQ^\theta}|\Phi_R}$. 
It is thus normal that it does not appear in Eq.~(\ref{vnappsyst}), which 
was derived by truncating the power-series expansion to order $z^2$.]

We now evaluate the systematic error on the differential flow. 
Our estimate of $v'_{mn}$ is defined by Eq.~(\ref{diffflow}). 
We assume for simplicity that the particle with angle $\psi$ 
is not involved in the definition of $Q^\theta$, i.e., we 
assume that autocorrelations have been removed. 

Since we neglect statistical errors, we can replace sample averages
by true averages (expectation values) in this formula. 
The denominator of the last factor in Eq.~(\ref{diffflow}) is 
\begin{eqnarray}
\label{bout1}
\mean{Q^\theta e^{z_0^\theta Q^\theta}}
&=&\frac{d G^\theta(z)}{d z}|_{z=z_0^\theta}\cr
&=& iMv_n(1-\epsilon)\,e^{M(1+v_{2n}-2v_n^2)(z_0^{\theta})^2/4}J_1(j_{01}).\cr
& & 
\end{eqnarray}
The numerator is evaluated by first taking the average value for 
a fixed $\Phi_R$, which is given by Eqs.~(\ref{average1}) and 
(\ref{shortrangeapp1}): 
\begin{widetext}
\begin{equation}
\mean{\cos[mn(\psi-\theta)]\,e^{z_0^\theta Q^\theta}|\Phi_R}=
v'_{mn}\,e^{M (1+v_{2n}-2v_n^2)(z_0^\theta)^2/4}
\cos(m\alpha)\,e^{Z\cos\alpha-\epsilon Z^2\sin^2\alpha},
\end{equation}
where $Z\equiv Mv_nz_0^\theta$. 
The average over $\Phi_R$ follows from Eq.~(\ref{lemme}) with the help 
of Eq.~(\ref{z0syst}):
\begin{equation}
\label{bout2}
\mean{\cos[mn(\psi-\theta)]\,e^{z_0^\theta Q^\theta}}=
i^m v'_{mn}\,e^{M (1+v_{2n}-2v_n^2)(z_0^\theta)^2/4}
\left(1+\epsilon m^2\right) J_m(j_{01}). 
\end{equation}
\end{widetext}
Using Eqs.~(\ref{diffflow}), (\ref{vnappsyst}), (\ref{bout1}), 
and (\ref{bout2}), we finally obtain 
\begin{equation}
\label{systvprime}
v^{\prime\theta}_{mn}\{\infty\} = v'_{mn}(1+\epsilon m^2).
\end{equation}
Therefore, the relative systematic error on $v'_{mn}$ is $\epsilon m^2$, 
where $\epsilon$ is defined in Eq.~(\ref{abreviations}). 
Note that the correction to the integrated flow, 
Eq.~(\ref{vnappsyst}), has the opposite sign.
In particular, integrating $v^{\prime\theta}_n\{\infty\}$
over phase space, one does not recover the integrated flow. 
This is because we have assumed that autocorrelations have 
been removed. 
Note also that the correction increases for higher harmonics.

Finally, note that Eqs.~(\ref{vnappsyst}) and (\ref{systvprime}) 
are quite general: they still hold if there are weights, or if 
nonflow correlations are present. The only difference is that 
$\epsilon$ is no longer given by Eq.~(\ref{abreviations})
(this equation remains valid as an order of magnitude, though).

\end{document}